\documentclass[preprint]{revtex4-1}

\usepackage[dvipsnames]{xcolor}

\usepackage{textcomp}

\usepackage{amssymb}
\usepackage{graphicx}
\usepackage{subfigure}
\usepackage{amsfonts}
\usepackage{amssymb}
\usepackage{latexsym}
\usepackage{natbib}
\usepackage{bm}
\usepackage{newlfont}
\usepackage{epsfig}
\usepackage{ctable}
\usepackage{amsmath}
\usepackage[american]{babel}

\usepackage{placeins}
\usepackage{cleveref}

\usepackage{lineno}

\crefname{section}{§}{§§}
\Crefname{section}{§}{§§}

\begin{document}

\title{\textcolor{black}{Lagrangian vs Eulerian view on the mean drift and streaming flows in orbital sloshing}} 

\author{A.~Bongarzone} 
\author{F.~Gallaire}\email{francois.gallaire@epfl.ch} 
\affiliation{Laboratory of Fluid Mechanics and Instabilities, \'Ecole Polytechnique F\'ed\'erale de Lausanne, Lausanne, CH-1015, Switzerland}

\begin{abstract}

\textcolor{black}{Orbital sloshing is a technique to gently mix a container's liquid content and it is commonly used in fermentation and cell cultivation processes. Besides the rich wave dynamics observed at the interface, Bouvard \textit{et al.} (2017) \cite{bouvard2017mean} unveiled the structure of the Lagrangian mean flow hiding in the fluid bulk. The latter flow shows a global toroidal (azimuthal) rotation co-directed with the wave and nontrivial poloidal vortices near the contact line. Rotating sloshing waves are known to induce a net motion of fluid particles and hence a wave-averaged difference between the Eulerian flow -- viscous streaming -- and the Lagrangian flow, that is commonly referred to as Stokes drift. Nevertheless, discerning these two components in an experiment is challenging as they scale similarly with the forcing amplitude and frequency. Their relative contributions remain therefore unquantified, particularly for highly viscous fluids, for which prior analysis, based on inviscid arguments, fails. In this work, we construct a truncated asymptotic approximation of the problem, where the solution at each order is computed numerically to describe as accurately as possible the contact line region, otherwise analytically untractable. The results of this weakly nonlinear analysis in terms of first-order wave and second-order mean flow are then thoroughly compared with the experiments by Bouvard \textit{et al.} (2017) \cite{bouvard2017mean}, showing a remarkable agreement for off-resonance frequencies. When viscosity matters, our findings suggest that it is incorrect to attribute the poloidal patterns solely to the Eulerian streaming flow and that viscous corrections to the Stokes drift are equally important in the resulting mean Lagrangian flow.}

\end{abstract}

\maketitle

\begin{centering}\section{Introduction}\label{sec:S1}\end{centering}

\textcolor{black}{Shaking a partially filled beaker is a standard technique in biological and chemical industrial applications including fermentation processes and drug production as well as bacterial and cellular cultures \citep{mcdaniel1969effect,wurm2004production}. Specifically, orbital shaking constitutes a simple and efficient way to prevent sedimentation and ensure homogenisation in the concentration of dissolved oxygen and nutrients. This can be achieved in both small and large-size bioreactors \citep{liu2001development,de2004tubespin,muller2007scalable} in a non-intrusive way, i.e. without any stirring, and possibly circumventing the problem of oxygen supply from the beaker's bottom, whose action is instead replaced by the gas exchange taking place at the liquid interface \citep{handa1989effect,kretzmer1991response,papoutsakis1991fluid}. The free surface deformations resulting from the imposed oscillatory motion cause a laminar and low-shear flow environment which is a requirement for optimal cellular growth \citep{kim2009stirring}. For all these reasons, a strong interest in the gas exchange and mixing processes taking place in these devices has emerged in the last few decades \citep{buchs2000power1,buchs2000power2,buchs2001introduction,maier2004advances,muller2005orbital,micheletti2006fluid,zhang2009efficient,tissot2010determination,tan2011measurement,tissot2011efficient,klockner2012advances}.\\
\indent At a pure hydrodynamic level, resonant sloshing dynamics in archetypical geometries such as circular cylindrical containers has progressively gained understanding thanks to a series of works which have combined experiments \citep{abramson1966dynamic,abramson1966some,chu1968subharmonic,royon2007liquid,hopfinger2009liquid,reclari2013hydrodynamics,reclari2014surface,bouvard2017mean,moisy2018counter} and theoretical modelling based on either multimodal theories \citep{faltinsen2016resonant,timokha2017,raynovskyy2018steady,raynovskyy2018damped,raynovskyy2020sloshing,horstmann2020linear,horstmann2021formation,chen2023mechanism} or multiple scale weakly nonlinear analyses \citep{bongarzone2022amplitude,marcotte2023super,marcotte2023swirling}. Those studies have elucidated the rich dynamics observed in close-to-resonance operating conditions and have provided powerful theoretical tools capable of predicting and explaining some of the physical mechanisms behind complex nonlinear phenomena such as large-amplitude super-harmonic resonances, symmetry-breaking conditions, counter-rotating waves, irregular and chaotic states, wave steepening and breaking conditions.\\
\indent Notwithstanding the body of literature on the sloshing hydrodynamics, most of the abovementioned theoretical studies have employed potential theories \citep{Lamb32,faltinsen2005liquid,ibrahim2009liquid} considering zero mean vorticity and therefore overlooking the hidden mean bulk motion associated with the wave dynamics and that it is key in the functioning principles of bioreactors or similar devices.\\
\indent Far from resonance, the steady-state response of a cylindrical container to a horizontal orbital motion at a constant angular frequency consists of a small-amplitude co-directed rotating wave that comes along with a mean flow \citep{hutton1963inv}. With their work focusing on the weakly nonlinear limit, where the amplitude of the waves and that of the associated mean flow are separated in their order of magnitude, Bouvard \textit{et al.} (2017) (B17) \cite{bouvard2017mean} have given important insights into the mechanism that generates the mean flow induced by orbital shaking and its dependence on the flow parameters, like the aspect ratio of the container, the fluid viscosity or the forcing amplitude. Particularly, they have unveiled and characterized the structure of the axisymmetric Lagrangian mean flow hidden in the fluid bulk, showing both a global toroidal \textcolor{black}{(or azimuthal in the $\theta$-direction)} rotation co-directed with the wave and nontrivial poloidal vortices near the contact line \textcolor{black}{(in the vertical $rz$-plane)}.\\
\indent In this regard, it is important to distinguish between the Eulerian and Lagrangian mean flow. The former is given by the time-averaged of the velocity field and it is commonly referred to as streaming flow. On the other hand, the Lagrangian mean flow corresponds to the flow that induces the total mean mass transport. The wave-averaged difference between the Eulerian and Lagrangian flow is commonly referred to as Stokes drift \citep{stokes1847theory,van2018stokes,vanneste2022stokes},}

\begin{equation}
\label{eq:EQ0}
\underbrace{\overline{\mathbf{u}}\left(r,z\right)}_{\text{Lagrangian}}=\underbrace{\overline{\mathbf{u}}^E\left(r,z\right)}_{\text{Eulerian}}+\underbrace{\overline{\mathbf{u}}^S\left(r,z\right)}_{\text{Stokes drift}}.
\end{equation}

\bigskip

\textcolor{black}{\indent When considering inviscid or nearly inviscid waves, the literature offers the following common interpretations for the two contributions appearing in this decomposition:\\
\indent (i) fluid particles in oscillatory flows follow nearly circular closed paths and undergo slight variations in wave magnitude along these trajectories \citep{bouvard2017mean}. These variations cause particles to displace more in one direction, resulting in a small average drift over each wave period \citep{van2018stokes}. Known as Stokes drift, this phenomenon originates purely from kinematics \citep{stokes1847theory} and enables \textcolor{black}{azimuthal} mass transport even without an Eulerian mean flow;\\
\indent \textcolor{black}{(ii) the Eulerian streaming flow has instead an intrinsically viscous origin as it represents the response of the flow to the time-averaged nonlinear Reynolds stress. As irrotational waves do not carry vorticity, the streaming flow stems from nonlinear interactions within the viscous boundaries layers at solid walls and free surface} \citep{batchelor1967introduction,craik1982drift,craik1976rational,perinet2017streaming,higuera2005dynamics,riley2001steady}.\\
\indent Unfortunately, the intensity of the streaming flow and Stokes drift obey the same scaling with the forcing amplitude and frequency, thus challenging the correct discernment of these two components in experiments performed using stroboscopic PIV as in B17. \\ 
\indent \textcolor{black}{For large ratios of liquid depth over radius, $h/R\gg1$,} Faltinsen \& Timokha (2019) (FT19) \cite{faltinsen2019inviscid} proposed an inviscid analytical theory describing a slow steady liquid mass rotation during the swirl-type sloshing in a vertical circular cylindrical tank. The theory uses the asymptotic steady-state wave solution by Faltinsen \textit{et al.} (2016) \cite{faltinsen2016resonant} and supplements the traditional inviscid hydrodynamic model with an inviscid Craik-Leibovich equation to describe the toroidal (azimuthal) part of the vortical Eulerian mean flow. Combining the latter with the Stokes drift contribution calculated directly from the wave kinematics, they computed the total azimuthal mean mass transport. Their analytical model compares remarkably well with prior experiments by Hutton (1963) \cite{hutton1964fluid} and, to some extent, with those by B17. Specifically, this inviscid model, as it is based on a multiple time-scale expansion of the problem close to the fundamental resonance frequency, is particularly suited for describing the near-resonance flow characteristics of inviscid waves, but its predictive power deteriorates far away from resonance, where most of the mean flow experiments by B17 were performed. Furthermore, the inviscid framework employed in FT19, although augmented with a Craik-Leibovich equation, overlooks the axisymmetric radial and axial components of the poloidal streaming flow, which are non-negligible when viscous effects matter, as in some of the cases considered by B17. \textcolor{black}{As we will see, these velocity components produce recirculating patterns in the vertical plane that are responsible for the top-to-bottom transport essential to achieve good mixing and promote gas exchange.}\\ 
\indent Predicting those poloidal structures inevitably requires adopting a viscous framework to correctly describe the effect of the boundary layer pumping on the fluid bulk and the resulting mean flow generation. In the context of standing Faraday waves, Périnet \textit{et al.} (2017) \cite{perinet2017streaming} considered a model for streaming flows in quasi-inviscid fluids that includes the complex coupling occurring in the viscous boundary layers. Through boundary layer matching, the coupling resulted in adjusted boundary conditions within a three-dimensional Navier–Stokes formulation of the streaming flow. Numerical simulations using this framework demonstrated satisfactory agreement, both qualitatively and quantitatively, with the velocity fields observed in their dedicated experiments.\\
\indent However, these authors did not focus on the complex contact line dynamics and how it might affect the streaming pattern, perhaps because the Hele-Shaw cells employed in their experimental campaign were such that contact line streaming became quickly unimportant sufficiently far from the lateral wall. This is in stark contrast with the sloshing experiments by B17, where the poloidal recirculation vortices are mostly active near the contact line and are fundamentally driven by the oscillatory boundary layers in the vicinity of this region.\\
\indent While multiple scale expansions and matched asymptotics can be successfully used to describe nonlinear interactions in regions near the rigid walls and free surface \citep{perinet2017streaming,nicolas2003three}, the contact line region, where the sloshing wave elevation is maximum, is even harder to consider as it is almost analytically intractable. This analytical difficulty is what has made predicting the structure of the Lagrangian mean flow in the orbital sloshing configuration such a challenging task so far.\\}
\textcolor{black}{\indent With the final goal of formalising a comprehensive framework capable of reproducing the experimental findings of B17 and predicting mean flow generation in a wider variety of configurations, the present study represents a first step in this direction. Here we construct a truncated asymptotic approximation of the problem, where the solution at each order is computed numerically to describe as accurately as possible the contact line region. Particular care is therefore devoted to the modelling of the solid wall and contact line boundary conditions. The results of this asymptotic expansion in terms of first-order wave and second-order mean flow are then thoroughly compared with the experiments by B17, showing a remarkable agreement, at least for off-resonance forcing frequencies. For highly viscous fluids, for which prior inviscid asymptotic models fail, our findings suggest that it is fundamentally wrong to attribute the poloidal patterns solely to the Eulerian streaming flow and that viscous corrections to the Stokes drift are equally important in the resulting Lagrangian flow.\\}
\textcolor{black}{\indent We point out that, in contradistinction to the analysis by FT19, which is designed to deal with resonant inviscid waves, we propose here a diametrically opposed approach \textcolor{black}{that can deal with non-resonant but viscous orbital sloshing and is, therefore, more suitable to compare with the experimental mean flow measurements by B17}. This study is therefore complementary to that of FT19 \citep{faltinsen2019inviscid}.\\}
\indent The manuscript is organized as follows. The considered flow configuration and governing equations, with their boundary conditions are given in \S\ref{sec:S2}. Section \S\ref{sec:S3} is dedicated to the viscous linear analysis. The resolution of the linear problem is carried out numerically and provides the first-order swirling sloshing wave, whose fundamental features are then compared with those experimentally quantified by B17. The characterization of the weakly nonlinear second-order Lagrangian mean flow and its decomposition into the Eulerian and Stokes drift components are tackled in \S\ref{sec:S4}. Final comments and conclusions are outlined in \S\ref{sec:S5}. Appendix~\ref{sec:A1} complements the theoretical modelling and the numerical analysis by computing the sensitivity of the mean flow to variations in the specifications of the sidewall boundary condition, whereas Appendix~\ref{sec:A2} extends the comparison with the experiments to other prior studies.

\begin{centering}\section{Flow modelling}\label{sec:S2}\end{centering}

\begin{figure}
\centering
\includegraphics[width=0.95\textwidth]{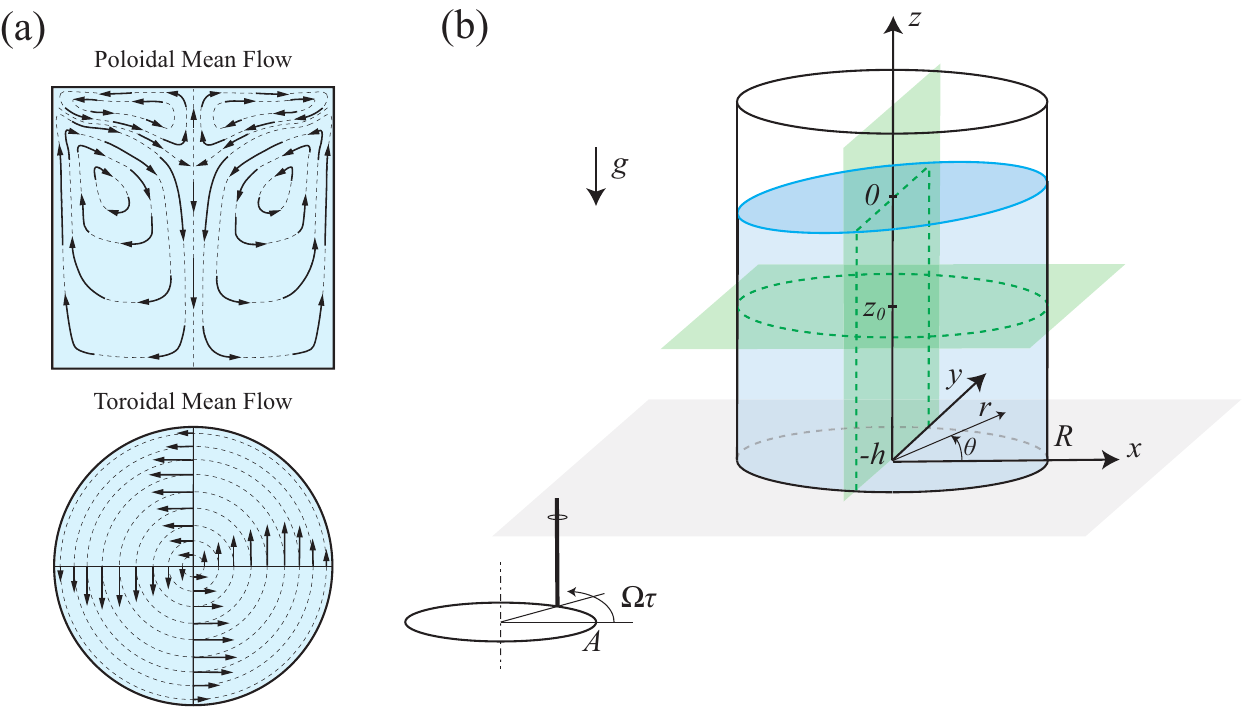}
\caption{(a) Cartoon representation of the poloidal and toroidal mean flow fields as described in Ref.~\onlinecite{bouvard2017mean} (B17). These two views correspond to the green-shaded planes in (b): the poloidal flow is represented in the vertical $yz$-plane, while the toroidal flow is shown in the horizontal $xy$-plane at a coordinate $z=z_0$ below the free surface. (b) Sketch of a cylindrical container of radius $R$ and filled to a depth $h$ with a liquid. The moving Cartesian reference frame is $O\hat{\mathbf{x}}\hat{\mathbf{y}}\hat{\mathbf{z}}$. The origin of the moving cylindrical reference frame $O\hat{\mathbf{r}}\hat{\boldsymbol{\theta}}\hat{\mathbf{z}}$ is placed at the container revolution axis and, specifically, at the unperturbed liquid height, $z=0$. The container keeps fixed its orientation but undergoes a time-harmonic circular orbit of radius $A$ and angular frequency $\Omega$.}
\label{fig:F1} 
\end{figure}

\begin{centering}\subsection{Geometry and shaking configuration}\label{subsec:S2s1}\end{centering}

Let us consider a vertical cylindrical container of radius $R$ filled to a depth $h$ with a liquid of density $\rho$ and dynamic viscosity $\mu$. The air-liquid surface tension is denoted by $\gamma$. A cylindrical coordinate system is defined, where $z$ is the vertical direction corresponding to the axis of the container, $r$ is the radial direction, $\theta$ is the azimuth, and the zero is set at the unperturbed interface position at the centerline as in figure~\ref{fig:F1}(b). The orbital (circular) shaking motion is represented as the combination of two sinusoidal translations with a $\pi/2$ phase shift, thus leading to the following equations of motion for the container axis intersection with the z = 0 plane, parametrized in cylindrical coordinates $\left(r,\theta\right)$,
\begin{equation}
\label{eq:EQ1}
\dot{\mathbf{x}}_c =
  \begin{cases}
    -A\Omega\sin{\left(\Omega \tau-\theta\right)}\,\hat{\mathbf{r}}\\
    \textcolor{white}{+}A\Omega\cos{\left(\Omega \tau-\theta\right)}\,\hat{\boldsymbol{\theta}}
  \end{cases}.
\end{equation}
\noindent with $A$ and $\Omega$ the dimensional forcing amplitude and angular frequency, respectively.

\begin{centering}\subsection{Governing equations}\label{subsec:S2s2}\end{centering}

\indent The fluid motion associated with a displacement of the interface, $\eta\left(r,\theta,t\right)$, is governed by the incompressible and unsteady Navier-Stokes equations:
\begin{equation}
\label{eq:EQ2}
\frac{\partial\mathbf{u}}{\partial t}+\nabla p-\frac{1}{Re}\Delta\mathbf{u}=-\left(\mathbf{u}\boldsymbol{\cdot}\nabla\right)\mathbf{u}+\epsilon\mathbf{f},\ \ \ \ \ \nabla\boldsymbol{\cdot}\mathbf{u}=0,
\end{equation}
\noindent which were made non-dimensional by using the container's characteristic length $R$ and forcing time scale $1/\Omega$. In~\eqref{eq:EQ2}, $\mathbf{u}\left(r,\theta,z,t\right)=\left\{u_r,u_{\theta},u_z\right\}^T$ denotes the velocity vector, $p\left(r,\theta,z,t\right)=P\left(r,\theta,z,t\right)+z/Fr$ is the reduced pressure and $P$ the total pressure. The non-dimensional external forcing reads 
\begin{equation}
\label{eq:EQ3}
\mathbf{f}=\left\{\cos{\left(t-\theta\right)},\sin\left(t-\theta\right),0\right\}^T,
\end{equation}
\noindent with $t=\Omega \tau$ the dimensionless time.\\
\indent At $z=\eta$, the kinematic boundary condition prescribes that no flow is allowed through the interface,
\begin{equation}
\label{eq:EQ4}
\frac{\partial\eta}{\partial t}-\left.u_z\right|_{\eta}=-\left.u_r\right|_{\eta}\frac{\partial\eta}{\partial r}-\frac{\left.u_{\theta}\right|_{\eta}}{r}\frac{\partial\eta}{\partial\theta},
\end{equation}
\noindent and the Laplace law gives the relation between the stress and the local curvature, $\kappa\left(\eta\right)$:
\begin{equation}
\label{eq:EQ5}
\left[-\left.p\right|_{\eta}+\frac{\eta}{Fr}+\frac{1}{Re}\left.\left(\nabla\mathbf{u}+\nabla^T\mathbf{u}\right)\right|_{\eta}\right]\mathbf{n}=\frac{1}{Bo\,Fr}\kappa\left(\eta\right)\mathbf{n}.
\end{equation}
\noindent with 
\begin{equation}
\label{eq:Eq5c}
\kappa\left(\eta\right)=\frac{\eta_{rr}\left(1+\frac{\eta_{\theta}^2}{r^2}\right)+\left(1+\eta_r^2\right)\left(\frac{\eta_r}{r}+\frac{\eta_{\theta\theta}}{r^2}\right)-2\eta_r\frac{\eta_{\theta}}{r^2}\left(\eta_{rr}+\frac{\eta_{\theta}}{r}\right)}{\left(1+\eta_r+\frac{\eta_{\theta}}{r}\right)^{3/2}}.
\end{equation}
\noindent The system is thus characterized by five nondimensional numbers:
\begin{eqnarray}
\label{eq:EQ6}
H=\frac{h}{R}=2.168,\ \ \ \ \epsilon=\frac{A}{R}\in\left[0.006,0.20\right],\ \ \ \ Fr=\frac{R\Omega^2}{g}\in\left[0.0135,0.45\right],\\
Bo=\frac{\rho gR^2}{\gamma}=1100,\ \ \ \ Re=\frac{\Omega R^2}{\nu}\in\left[50,1500\right],\ \ \ \ \ \ \ \ \ \ \ \ \ \ \ \ \ \nonumber
\end{eqnarray}
\noindent in order, the aspect ratio of the cylinder $H$, the normalized forcing amplitude $\epsilon$, the Bond number $Bo$, the Reynolds number $Re$, and the Froude number $Fr$. Note that the Bond and Reynolds numbers here defined give, respectively, non-dimensional measures of the capillary length $l_{cap}=\sqrt{\gamma/\rho g}$ and the Stokes boundary layer thickness $l_{St}=\sqrt{2\nu/\Omega}$. The values given in~\eqref{eq:EQ6} correspond to the range of parameters investigated in Ref.~\onlinecite{bouvard2017mean} (B17), to which we will restrict our analysis. These values correspond to the case of a circular cylindrical container of radius $R=51.2\,\text{mm}$ filled to a depth $h=111\,\text{mm}$ with two different silicon oils of air-liquid surface tension $\gamma=21\times10^{-3}\,\text{N\,m$^{-1}$}$ and kinematic viscosity $\nu=50$ or $500\,\text{mm$^2$\,s$^{-1}$}$.

\bigskip

\begin{centering}\subsection{Treatment of the sidewall: a macroscopic depth-dependent slip-length model}\label{subsec:S2s3}\end{centering}

While the no-slip and no-penetration boundary conditions represent natural choices for the modelization of the solid bottom, the modelling of the sidewall boundary condition is more subtle and requires different strategies. It is indeed well-assessed in the literature that the no-slip condition and a moving contact line are not compatible with each other \citep{Huh71,Davis1974} as their simultaneous imposition would yield a stress singularity at the moving contact line. A common treatment consists in adopting a slip-length model, thus assuming that the fluid speed relative to the solid wall is proportional to the viscous stress \citep{navier1823memoire,Lauga2007,Viola2018b} and that, together with the no-penetration condition, provides the boundary conditions
\begin{equation}
\label{eq:EQ7}
u_r=0,\ \ \ \ \ \ u_{\theta}=l_s\left(z\right)\left(\frac{\partial u_{\theta}}{\partial r}-u_{\theta}\right),\ \ \ \ \ \ u_z=l_s\left(z\right)\frac{\partial u_z}{\partial r},\ \ \ \ \ \ \text{at $r=1$}.
\end{equation}
\noindent It was hypothesized in Refs.~\onlinecite{miles1990capillary} and~\onlinecite{ting1995boundary} that the phenomenological macroscopic slip length appearing in equation~\eqref{eq:EQ7} should not be assumed constant along the wall, but rather spatially dependent on the position along the lateral wall and vanishing at a certain distance away from the contact line, where the flow obeys the no-slip condition. With this in mind, we employ here a depth-dependent slip length as proposed in Refs.~\onlinecite{bongarzone2022numerical,bongarzone2024stick}, which has been shown to correctly estimate the linear dissipation originating in the Stokes boundary layers at the lateral solid walls in various flow configurations. Briefly, we postulate that the slip length $l_s\left(z\right)$ is described by the exponential law
\begin{equation}
\label{eq:EQ8}
l_s\left(z\right)=l_{cl}\,\text{exp}\left(-\frac{z}{\delta}\text{log}\left(\frac{l_{\delta}}{l_{cl}}\right)\right),\ \ \ \ z\in\left[-H,0\right].
\end{equation}
\noindent In equation~\eqref{eq:EQ8}, $l_{cl}$ is the slip-length value at the contact line, whereas $l_{\delta}$ is its value at a distance $\delta$ below the contact line, $z=-\delta$, with $\delta$ representing the size of the slip region \citep{ting1995boundary}. In principle, $l_{cl}$, $l_{\delta}$ and $\delta$ are all free parameters. However, keeping in mind that, macroscopically speaking, we aim at mimicking a stress-free condition in the vicinity of the contact line, as we assume the latter to obey the free-end edge condition,
\begin{equation}
\label{eq:EQcl}
\left.\frac{\partial\eta}{\partial r}\right|_{r=1}=0,
\end{equation}
\noindent and at reproducing the no-slip condition after a certain distance $\delta$, a natural choice consists in using a penalization-like technique where a large value of $l_{cl}\gg1$ ($\sim 10^{2}$$\div$$10^{4}$) and a small value of $l_{\delta}\ll1$ ($\sim 10^{-4}$$\div$$10^{-6}$) are imposed. The range of values proposed in brackets is based on the sensitivity analysis reported in Ref.~\onlinecite{bongarzone2022numerical}. The remaining free parameter is the slip region penetration depth, $\delta$. What mostly matters in choosing the value of this parameter is that $\delta$ should be kept small compared to all the other physical scales at hand in the problem, i.e. $H$, $R$, capillary length $1/\sqrt{Bo}$, Stokes boundary layer thickness $\sqrt{2/Re}$ or, at most, comparable to the smallest one. Hence, results presented throughout the manuscript are obtained by fixing $\delta=1/\sqrt{Bo}\approx0.03$ (only 1.4\% of H), while the effect of this parameter on the flow fields is more thoroughly explored in Appendix~\ref{sec:A1} confirming the need to keep delta small enough.

\bigskip

\begin{centering}\section{Viscous linear analysis: first-order swirling sloshing wave}\label{sec:S3}\end{centering}

Sufficiently far from fundamental harmonic resonances or any other secondary resonances \citep{bongarzone2022amplitude,marcotte2023super} and in the limit of small forcing amplitude $\epsilon=A/R\ll1$, the linear viscous analysis provides a good approximation of the time-harmonic wave response. Let us first decompose asymptotically the flow fields as
\begin{equation}
\label{eq:EQasy}
\mathbf{q}\left(r,\theta,z,t\right)=\left\{\mathbf{u},p,\eta\right\}^T=\epsilon\mathbf{q}'\left(r,\theta,z,t\right)+\epsilon^2\overline{\mathbf{q}}^E\left(r,z\right)+\mathcal{O}\left(\epsilon^3\right),
\end{equation}
\noindent where the first term, $\epsilon\mathbf{q}'$, represents the linear time-harmonic sloshing swirling wave, whereas the second term, $\epsilon^2\overline{\mathbf{q}}^E$, indicates the associated second-order time-averaged axisymmetric Eulerian mean flow, whose characterization constitutes one of the main purposes of this work and will be tackled in \S\ref{sec:S4}\\
\indent At order $\epsilon$, governing equations reduce to a forced Stokes problem supplemented by the kinematic equation describing the evolution of the interface linear perturbation. This problem can be conveniently recast in an operator form as 
\begin{equation}
\label{eq:EQ9}
\left(\partial_t\mathcal{B}-\mathcal{A}\right)\mathbf{q}'=\mathbf{f},
\end{equation}
\noindent with 
\begin{equation}
\label{eq:EQ10}
\mathcal{B}=\left(
\begin{matrix}
 I_{\mathbf{u}} & 0 & 0\\
0 & 0 & 0\\ 
0 & 0 & I_{\eta}\\
\end{matrix}
\right), \ \ \ \ \ 
\mathcal{A}=\left(
\begin{matrix}
Re^{-1}\Delta & -\nabla & 0\\
\nabla^T & 0 & 0\\ 
I_{u_z} & 0 & 0\\
\end{matrix}
\right), \ \ \ \ \ 
\mathbf{q}'=\left(
\begin{matrix}
\mathbf{u}'\\
p'\\
\eta'
\end{matrix}
\right),
\end{equation}
\noindent and where $I_{\mathbf{u}}$, $I_{u_z}$ and $I_{\eta}$ are, respectively, the identity matrices associated with the velocity field $\mathbf{u}$, the axis velocity component $u_z$ and the interface $\eta$. Note that the components of the vectorial stress condition~\eqref{eq:EQ5} linearized around the rest state with a flat static interface, i.e.
\begin{subequations}
\begin{equation}
\label{eq:EQstresstan}
\left.\frac{\partial u_r'}{\partial z}\right|_0+\left.\frac{\partial u_z'}{\partial r}\right|_0=0,\ \ \ \ \ \ \ \left.\frac{\partial u_{\theta}'}{\partial z}\right|_0+\frac{1}{r}\left.\frac{\partial u_z'}{\partial \theta}\right|_0=0,
\end{equation}
\begin{equation}
\label{eq:EQstressnorm}
-\left.p'\right|_{0}+\frac{\eta'}{Fr}+\frac{2}{Re}\left.\frac{\partial u_z'}{\partial z}\right|_{0}-\frac{1}{Bo\,Fr}\left.\frac{\partial\kappa\left(\eta\right)}{\partial \eta}\right|_0\eta'=0,
\end{equation}
\end{subequations}
\noindent do not explicitly appear in~\eqref{eq:EQ9}-\eqref{eq:EQ10}, but they are rather enforced as a boundary condition at the interface as in Refs.~\onlinecite{bongarzone2021relaxation} and~\onlinecite{bongarzone2022sub}. The same holds for the Navier slip condition~\eqref{eq:EQ7} as well as for the no-slip conditions at the bottom and the free-end edge contact line condition~\eqref{eq:EQcl}, which are enforced at the lateral wall, at the solid bottom and the contact line, respectively. Given the time and azimuthal periodicity of the external forcing~\eqref{eq:EQ3}
\begin{equation}
\label{eq:EQ11}
\mathbf{f}=\left\{1/2,-\text{i}/2,0\right\}^Te^{\text{i}\left(t-\theta\right)}+c.c.=\hat{\mathbf{f}}e^{\text{i}\left(t-\theta\right)}+c.c.\,,
\end{equation}
\noindent one can seek a solution to system~\eqref{eq:EQ9} obeying the normal mode ansatz
\begin{equation}
\label{eq:EQ12}
\mathbf{q}\left(r,\theta,z,t\right)=\hat{\mathbf{q}}\left(r,z\right)e^{\text{i}\left(t-\theta\right)}+c.c.\,,
\end{equation}
\noindent where c.c. stands for complex conjugate. Substitution of~\eqref{eq:EQ11} and~\eqref{eq:EQ12} eventually leads to the following linear problem
\begin{equation}
\label{eq:EQ13}
\left(\text{i}\mathcal{B}-\mathcal{A}_{-1}\right)\hat{\mathbf{q}}=\hat{\mathbf{f}}.
\end{equation}
\noindent We note that, after introducing the azimuthal decomposition into wavenumbers $m$, the linear operator $\mathcal{A}$ depends on this wavenumber $m$, that is why a subscript $-1\,\left(=m\right)$ has been introduced in~\eqref{eq:EQ13}. The operator $\mathcal{A}$ also implicitely depends on the driving frequency $\Omega$ through the Reynolds number definition given in~\eqref{eq:EQ6}.

\bigskip

\begin{centering}\subsection{Discretization technique}\label{subsec:S3s1}\end{centering}

The linear operators $\mathcal{B}$ and $\mathcal{A}_{-1}$ are discretized in space by means of a staggered Chebyshev--Chebyshev collocation method implemented in Matlab analogous to that described in Ref.~\onlinecite{bongarzone2022sub}. The three velocity components are discretized using a Gauss--Lobatto--Chebyshev (GLC) grid, whereas the pressure is staggered on Gauss--Chebyshev (GC) grid. Accordingly, the momentum equation is collocated at the GLC nodes and the pressure is interpolated from the GC to the GLC grid, while the continuity equation is collocated at the GC nodes and the velocity components are interpolated from the GLC to the GC grid. This results in the classical $P_N$-$P_{N-2}$ formulation, which is sufficient to suppress spurious pressure modes in the discretized problem \citep{Viola2016a,Viola2018b}. A two-dimensional mapping is then used to map the computational space onto the physical space. Lastly, the partial derivatives in the computational space are mapped onto the derivatives in the physical space, which depend on the mapping function. For further details see Refs.~\onlinecite{heinrichs2004spectral,canuto2007spectral,sommariva2013fast,Viola2018a}.\\
\indent Mesh convergence was tested for different refinements, starting from a grid size of $N_r=40$ and $N_z=40$ up to $N_r=100$ and $N_z=160$ with a progressive increment of $5$ GLC nodes along directions $r$ and $10$ GLC nodes along $z$. $N_r$ and $N_z$ denote here the number of radial and axial nodes, respectively. A mesh size of $N_r=75$ and $N_z=110$ was seen to be sufficient to ensure a good convergence of the results that will be discussed in the following. Particularly, a Chebyshev grid made of 110 points in the axial direction contains 10 grid points along the lateral wall within a distance $\delta=1/\sqrt{Bo}=0.03$ from the interface.\\
\indent After discretization, equation~\eqref{eq:EQ13} rewrites
\begin{equation}
\label{eq:EQ14}
\left(\text{i}\mathbf{B}-\mathbf{A}_{-1}\right)\hat{\mathbf{q}}=\hat{\mathbf{f}}.
\end{equation}
\noindent and, at least in non-resonant conditions, i.e. when $\Omega$ does not coincide with one of the natural frequencies of the system, can be straightforwardly resolved in Matlab for an imposed driving frequency $\Omega$.\\
\indent Lastly, we note that the numerical scheme requires explicit boundary conditions at $r=0$ to regularize the problem on the revolution axis \citep{Viola2018b}:
\begin{subequations}
\begin{equation}
\label{eq:EQ15a}
m=0:\ \hat{u}_r=\hat{u}_{\theta}=\frac{\partial\hat{u}_z}{\partial r}=\frac{\partial\hat{p}}{\partial r}=\frac{\partial\hat{\eta}}{\partial r}=0,
\end{equation}
\begin{equation}
\label{eq:EQ15b}
|m|=1:\ \frac{\partial\hat{u}_r}{\partial r}=\frac{\partial\hat{u}_{\theta}}{\partial r}=\hat{u}_z=\hat{p}=\hat{\eta}=0,\ \ 
\end{equation}
\begin{equation}
\label{eq:EQ15c}
|m|>1:\ \hat{u}_r=\hat{u}_{\theta}=\hat{u}_z=\hat{p}=\hat{\eta}=0.\ \ \ \ \ \ \
\end{equation}
\end{subequations}

\begin{figure}
\centering
\includegraphics[width=0.8\textwidth]{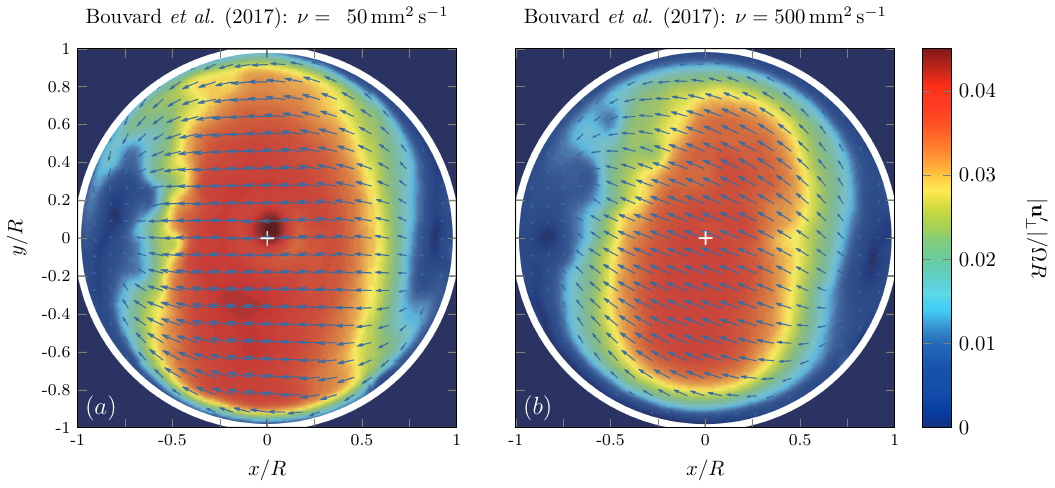}
\includegraphics[width=0.8\textwidth]{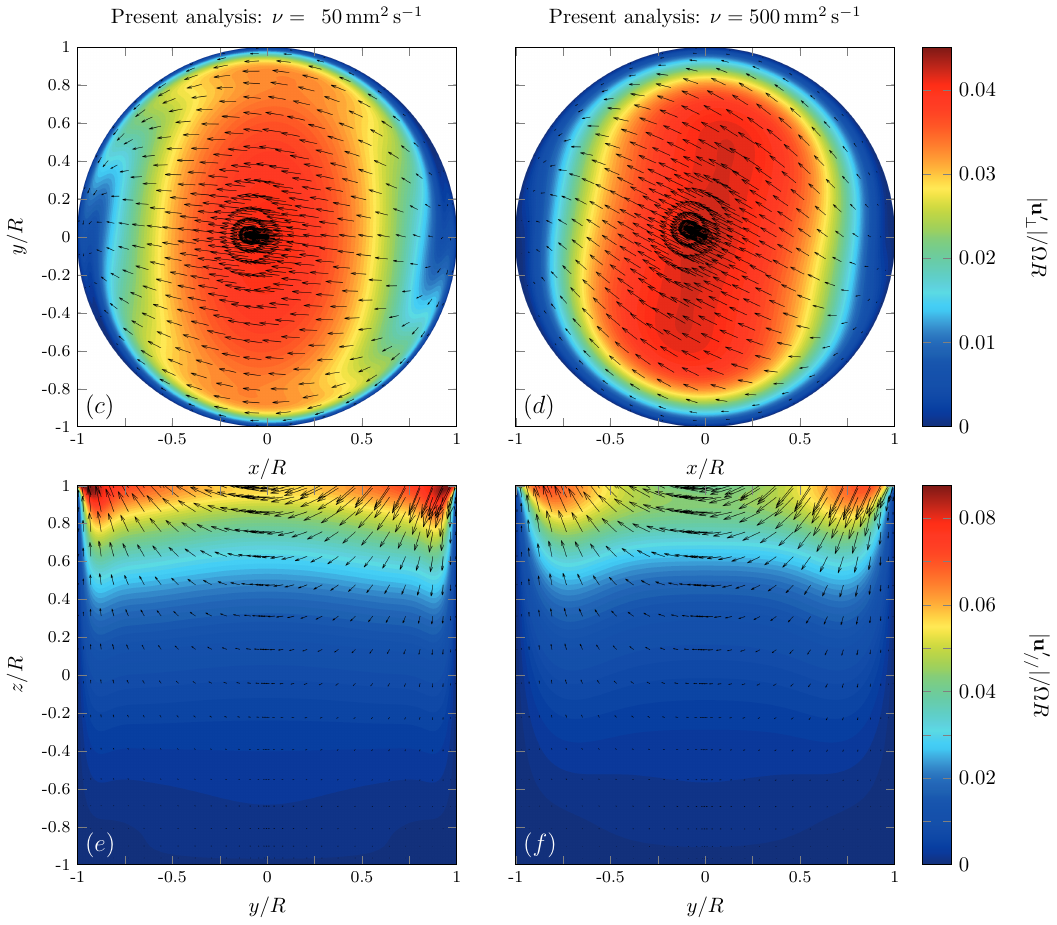}
\caption{Instantaneous snapshots of the sloshing wave taken at a phase $\pi/2$, for which the cylinder velocity is aligned with the unit vector $-\hat{\mathbf{x}}$. (a)-(b) Horizontal wave velocity experimentally measured by B17 at a driving frequency $\Omega/\omega_{1}=0.67$, amplitude $\epsilon=0.057$, and for a kinematic viscosity $\nu=50$ (a) and $500\,\text{mm$^2$\,s$^{-1}$}$ (b). The snapshots are taken at a coordinate $z=z_0=-0.23$ below the free surface. The colour map represents the rescaled velocity magnitude $|\mathbf{u}_{\perp}'|$, with $\epsilon u_r'$ and $\epsilon u_{\theta}'$ given by the arrows. (c)-(d) Same as in (a)-(b) but numerically computed according to~\eqref{eq:EQ14}. (e)-(f) Vertical wave velocity $|\mathbf{u}_{//}'|$ associated with snapshots (a)-(b).}
\label{fig:F2} 
\end{figure}

\bigskip

\begin{centering}\subsection{Comparison with experiments by Bouvard \textit{et al.} (2017)}\label{subsec:S3s2}\end{centering}

\begin{figure}
\centering 
\includegraphics[width=\textwidth]{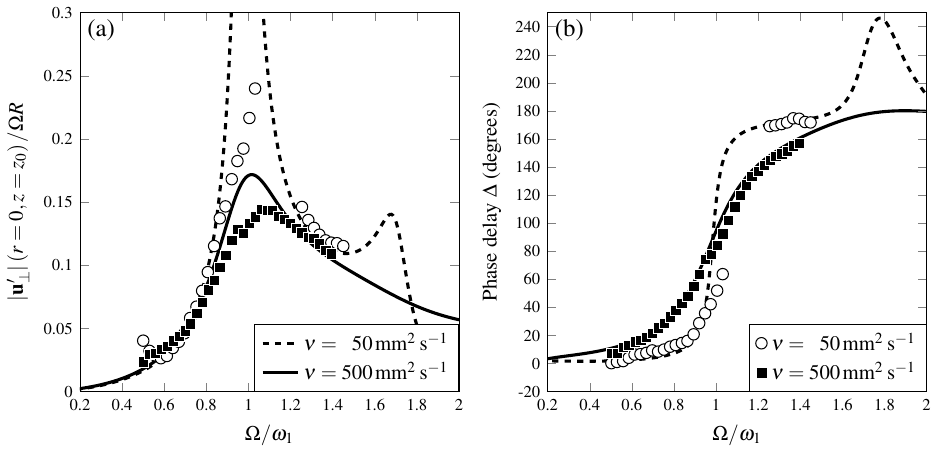}
\includegraphics[width=0.5\textwidth]{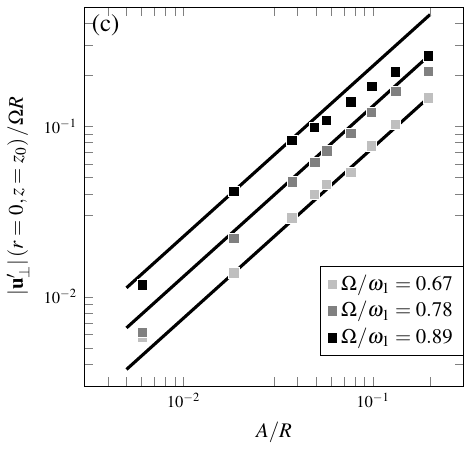}
\caption{This figure aims to reproduce figure~3 of B17. (a) Norm of the horizontal wave velocity measured at $r=0$, $z=z_0=-0.23$ , as a function of the forcing frequency $\Omega/\omega_{1}$ at a fixed forcing amplitude $\epsilon=0.057$, for the two different fluid viscosities. (b) Phase delay $\Delta$ between the forcing and the wave field at $r=0$ and defined as the angle between the fluid velocity at the centre of the cylinder and the unit vector $-\hat{\mathbf{x}}$. (c) Norm of the wave velocity as a function of the forcing amplitude $\epsilon=A/R$ at three values of the forcing frequency, for $\nu=500\,\text{mm$^2$\,s$^{-1}$}$. Markers correspond to the experiments by B17, whereas black solid and dashed lines show the present predictions according to the viscous linear solution of~\eqref{eq:EQ14}.}
\label{fig:F3} 
\end{figure}

In this section, we compare the prediction from the present numerically-based viscous linear analysis with experiments by B17. This comparison is outlined in terms of two fundamental flow quantities such as the wave fields in the horizontal, $\mathbf{u}_{\perp}'$, and vertical planes, $\mathbf{u}_{//}'$,
\begin{equation}
\label{eq:EQhv}
\frac{|\mathbf{u}_{\perp}'|\left(r,\theta,z,t\right)}{\Omega R}=\epsilon\sqrt{\mathbf{u}_r'^2+\mathbf{u}_{\theta}'^2},\ \ \ \ \ \ \ \ \ \ \ \frac{|\mathbf{u}_{//}'|\left(r,\theta,z,t\right)}{\Omega R}=\epsilon\sqrt{\mathbf{u}_r'^2+\mathbf{u}_z'^2},
\end{equation}
\noindent and phase shifts with the external forcing. Figure~\ref{fig:F2} aims precisely to reproduce their figure~2, showing instantaneous snapshots of the sloshing wave taken at a phase $\pi/2$, for which the cylinder velocity is aligned with the unit vector $-\hat{\mathbf{x}}$. Panels (a) and (b) show the rescaled horizontal wave velocity experimentally measured at a driving frequency $\Omega/\omega_{1}=0.67$, forcing amplitude $\epsilon=0.057$ ($A=2.9\,\text{mm}$), and for a kinematic viscosity $\nu=50$ ($Re=660$) and $500\,\text{mm$^2$\,s$^{-1}$}$ ($Re=66$), respectively. Although we do not neglect surface tension in our analysis, for the sake of comparison with B17, the value of the lowest system's natural frequency $\omega_1$ used to rescale the driving frequency is taken from the dispersion relation of pure gravity waves \citep{Lamb32},
\begin{equation}
\label{eq:EQ16}
\omega_n^2=k_n\tanh{\left(k_nH\right)}\ \ \ \ \ \ \text{with}\ \ \ \ \ \ n=1,2,\hdots\ \ \ \ \ \ \left(m=\pm 1\right),
\end{equation}
\noindent where the wavenumbers $k_n$ are the $n$th zeros of the derivative of the Bessel function of the first kind and first order $J_1$ ($k_1\simeq 1.841$, $k_2\simeq 5.331$, $\hdots$). \textcolor{black}{In the configurations here considered, the frequency correction to~\eqref{eq:EQ16} due to surface tension is approximately 0.1\% and therefore negligible.}\\
\indent The numerical wave fields reported in panels (c) and (d) are in good qualitative and quantitative agreement with the experimental ones shown in panels (a) and (b). Specifically, the boundary layers at the cylinder wall and the phase delay, more significant for the larger viscosity case, are both very well captured by the numerical analysis.\\
\indent Figure~\ref{fig:F3} compares the amplitude of the wave field, defined as the norm of the horizontal velocity at the centre, and its phase delay as systematically measured by B17 for driving frequencies $\Omega/\omega_{1}\in\left[0.5,1.5\right]$ and at a fixed forcing amplitude $\epsilon=0.057$. For the lower viscosity, the measured wave amplitude diverges as expected around $\Omega/\omega_{1} \approx 1$ and the resonance curve displays a hardening-like behaviour with a clear hysteresis, which is a typical nonlinear feature and is overlooked by the present linear model \citep{reclari2014surface,faltinsen2016resonant,marcotte2023swirling}. We anticipate that several approaches, some of which are mentioned in the final comments \S\ref{sec:S5}, could be employed to account for nonlinear effects. However, this is not the main purpose of this work.
\\
\indent With regards to the more viscous fluid, viscous linear effects are sufficient to smoothen the resonance curve, which shows a low amplitude peak shifted at $\Omega/\omega_{1} \approx 1.1$. \textcolor{black}{The computed peak occurs around $\Omega/\omega_{1}\approx1.01$, i.e. for $\Omega>\omega_{1}$, despite the presence of viscous dissipation. As explained in Ref.~\onlinecite{horstmann2020linear}, this counterintuitive behaviour is caused by the dependence of the forcing amplitude on the driving frequency. Nevertheless, such a shift in the resonant frequency is not sufficient to explain a 10\% increase, thus suggesting that nonlinear hardening-type effects are to some extent at play.} The phase delay between the forcing and the wave velocity, defined as the angle between the fluid velocity at the centre of the cylinder and $-\hat{\mathbf{x}}$, and shown in figure~\ref{fig:F3}(b), is well captured by the present model even for the less viscous fluid, except for the hysteretic jump at $\Omega/\omega_{1} \approx 1.1$.\\
\indent A further comparison is performed for variation of the forcing amplitude $\epsilon$ at fixed driving frequencies $\Omega/\omega_{1}$. This is reported in figure~\ref{fig:F3}(c), showing also a good agreement between the viscous numerical analysis and experiments for $\Omega/\omega_{1}=0.67$ and $\Omega/\omega_{1}=0.78$, whereas the agreement starts to deteriorate by increasing $\epsilon$ closer to the resonant frequency, e.g. at $\Omega/\omega_{1}=0.89$.\\
\indent For the sake of clarity, we did not include in figure~\ref{fig:F3} the curves from the inviscid potential model of B17, but the improvement is striking, especially for the more viscous fluid, for which dampening effects cannot be neglected.\\
\indent As a final comment to this section, we note the presence of a second resonant peak in figure~\ref{fig:F3}(a) and (b), which corresponds to the second fundamental natural frequency of the system, i.e. $\omega_2$. No experiments were reported for driving frequencies larger than $\Omega/\omega_{1}=1.5$, but the presence of this peak and its maximum value in panel (a) is consistent with that predicted by the linear damped interfacial wave theory proposed in Refs.~\onlinecite{horstmann2020linear,horstmann2021formation}. These authors also compare their predictions with those from B17. By looking at their figure~6, in comparison to figure~\ref{fig:F3} of this manuscript, it is not evident whether their analysis or the present one is in closer agreement with experiments, as they both give very similar estimates of the underlying flow quantities. Nevertheless, the substantial difference consists in the fact that in Refs.~\onlinecite{horstmann2020linear,horstmann2021formation} the wave field is constructed in the potential limit and only englobes viscous effects in the form of linear damping coefficient. On the contrary, in this work, we numerically solve the actual viscous problem to recover the details of flow fields, i.e. the boundary layers visible in figure~\ref{fig:F2}(c)-(f).\\
\indent This constitutes a main contribution of this work, as the computation of the correct wave field and, specifically, of the boundary layers and the corner region near the contact line, are essential in the prediction of the second-order Lagrangian mean flows tackled in the following section.


\bigskip

\begin{centering}\section{Viscous weakly nonlinear analysis: second-order Lagrangian mean flow}\label{sec:S4}\end{centering}

Recalling the definition given in \S\ref{sec:S1}, the total Lagrangian mean flow, $\overline{\mathbf{u}}^L$, is given by the sum of the Eulerian mean flow, $\overline{\mathbf{u}}^E$, and the Stokes drift, $\overline{\mathbf{u}}^S$:
\begin{equation}
\label{eq:EQ17}
\overline{\mathbf{u}}\left(r,z\right)=\overline{\mathbf{u}}^E\left(r,z\right)+\overline{\mathbf{u}}^S\left(r,z\right).
\end{equation}
\noindent In their campaign, Bouvard \textit{et al.} (2017) (B17) \cite{bouvard2017mean} only characterized the total Lagrangian mean flow, but could not experimentally separate the Eulerian and Stokes drift contributions. Using the numerical tools developed in \S\ref{sec:S3}, in this section we will first predict $\overline{\mathbf{u}}$ and compare it to experiments. Successively, we will provide a thorough characterization of the two individual contributions, $\overline{\mathbf{u}}^E$ and $\overline{\mathbf{u}}^S$.

\bigskip

\begin{centering}\subsection{Eulearian streaming flow}\label{subsec:S4s1}\end{centering}

In the weak steady streaming limit, the streaming flow can be computed asymptotically according to~\eqref{eq:EQasy} as a second-order correction to the first-order wave field. By introducing the (non-dimensional) time-average $\langle\mathbf{v}\rangle=\left(2\pi\right)^{-1}\int_0^{2\pi}\mathbf{v}\,\text{d}t$, with $\mathbf{v}$ a generic vectorial field, at order $\epsilon^2$ the time-averaged bulk equations reduce to a steady Stokes problem linearly forced by nonlinear terms which are quadratic in the first-order wave solution:
\begin{equation}
\label{eq:EQ18}
\nabla\overline{p}^E-\frac{1}{Re}\Delta\overline{\mathbf{u}}^E=\langle\left(\mathbf{u}'\boldsymbol{\cdot}\nabla\right)\mathbf{u}'\rangle,\ \ \ \ \ \nabla\boldsymbol{\cdot}\overline{\mathbf{u}}^E=0,
\end{equation}
\noindent subjected to the no-slip boundary condition at the solid bottom, $z=-H$, the Navier-slip condition at the lateral wall, $r=1$, the free-end edge contact line condition at $r=1$ and $z=0$, and to the steadily forced kinematic condition at the interface, $z=0$,
\begin{equation}
\label{eq:EQ19}
-\left.\overline{u}_z^E\right|_0=\langle\eta'\left.\frac{\partial u_z'}{\partial z}\right|_0-\left.u_r'\right|_0\frac{\partial \eta'}{\partial r}-\frac{1}{r}\left.u_{\theta}'\right|_0\frac{\partial\eta'}{\partial \theta}\rangle,
\end{equation}
\noindent where also the three dynamic equations are imposed:
\begin{subequations}
\begin{eqnarray}
\label{eq:EQ20a}
\frac{1}{Re}\left.\left(\frac{\partial \overline{u}_r^E}{\partial z}+\frac{\partial \overline{u}_z^E}{\partial r}\right)\right|_0=\ \ \ \ \ \ \ \ \ \ \ \ \ \ \ \ \ \ \ \ \ \ \ \ \ \ \ \ \ \ \ \ \ \ \ \ \ \ \ \ \ \\
\textcolor{black}{\langle n_r'\left(-\mathcal{P}'+\frac{2}{Re}\left.\frac{\partial u_r'}{\partial r}\right|_0\right)+\frac{n_{\theta}'}{Re}\left.\left(r\frac{\partial}{r}\left(\frac{u_{\theta}'}{r}\right)+\frac{1}{r}\frac{\partial u_r'}{\partial\theta}\right)\right|_0-\frac{\eta'}{Re}\frac{\partial}{\partial z}\left.\left(\frac{\partial u_r'}{\partial z}+\frac{\partial u_z'}{\partial r}\right)\right|_0\rangle},\notag
\end{eqnarray}
\begin{eqnarray}
\label{eq:EQ20b}
\frac{1}{Re}\left.\left(\frac{\partial \overline{u}_{\theta}^E}{\partial z}+\frac{1}{r}\frac{\partial \overline{u}_z^E}{\partial \theta}\right)\right|_0=\ \ \ \ \ \ \ \ \ \ \ \ \ \ \ \ \ \ \ \ \ \ \ \ \ \ \ \ \ \ \ \ \ \ \ \ \ \ \ \ \ \ \ \ \ \ \ \ \ \ \ \ \\
\textcolor{black}{\langle\frac{n_r'}{Re}\left.\left(r\frac{\partial}{r}\left(\frac{u_{\theta}'}{r}\right)+\frac{1}{r}\frac{\partial u_r'}{\partial\theta}\right)\right|_0+n_{\theta}'\left(-\mathcal{P}'+\frac{2}{Re}\left.\left(\frac{1}{r}\frac{\partial u_{\theta}'}{\partial \theta}+\frac{u_r'}{r}\right)\right|_0\right)-\frac{\eta'}{Re}\frac{\partial}{\partial z}\left.\left(\frac{\partial u_{\theta}'}{\partial z}+\frac{1}{r}\frac{\partial u_z'}{\partial \theta}\right)\right|_0\rangle},\notag
\end{eqnarray}
\begin{eqnarray}
\label{eq:EQ20c}
-\left.\overline{p}^E\right|_{0}+\frac{\overline{\eta}^E}{Fr}+\frac{2}{Re}\left.\frac{\partial \overline{u}_z^E}{\partial z}\right|_{0}-\frac{1}{Bo\,Fr}\left.\frac{\partial\kappa\left(\eta\right)}{\partial \eta}\right|_0\overline{\eta}^E=\ \ \ \ \ \ \ \ \ \ \ \ \ \ \ \ \\
\textcolor{black}{\langle\frac{n_r'}{Re}\left.\left(\frac{\partial u_r'}{\partial z}+\frac{\partial u_z'}{\partial r}\right)\right|_0+\frac{n_{\theta}'}{Re}\left.\left(\frac{\partial u_{\theta}'}{\partial z}+\frac{1}{r}\frac{\partial u_z'}{\partial \theta}\right)\right|_0-\eta'\frac{\partial}{\partial z}\left.\left(-p'+\frac{2}{Re}\frac{\partial u_z'}{\partial z}\right)\right|_0\rangle}.\notag
\end{eqnarray}
\end{subequations}
\noindent with $\mathbf{n}'=\left\{n_r',n_{\theta}',n_z'\right\}^T=\left\{-\partial_r\eta',-r^{-1}\partial_{\theta}\eta',1\right\}^T$ and $\mathcal{P}'=\left.p'\right|_0-\frac{\eta'}{Fr}+\frac{1}{Bo\,Fr}\left.\frac{\partial\kappa\left(\eta\right)}{\partial\eta}\right|_0\eta'$.\\
\indent Given the time- and azimuthal-periodicity of the first-order wave solution~\eqref{eq:EQ12}, time-averaging is equivalent to azimuthally-averaging, hence implying that the forcing terms appearing in~\eqref{eq:EQ18}-\eqref{eq:EQ20c} are altogether axisymmetric and will therefore induce an axisymmetric streaming flow, i.e. with $m=0$. It is important to note that this holds only for the case of a purely circular orbit, for which the forcing writes as in equation~\eqref{eq:EQ11}, while it is generally not the case when considering elliptic shaking conditions \citep{faltinsen2016resonant,marcotte2023swirling}.\\
\indent \textcolor{black}{After discretisation, the second-order problem is compactly written as}
\begin{equation}
\label{eq:EQ22}
-\mathbf{A}_{0}\overline{\mathbf{q}}^E=\langle\hat{\boldsymbol{F}}\rangle.
\end{equation}
\noindent that, assuming the non-singularity of the discretized operator $\mathbf{A}_{0}$, is straightforwardly resolved in Matlab giving in input the forcing vector $\hat{\boldsymbol{F}}$, which is only a function of the first-order wave solution computed in \S\ref{sec:S3}.

\bigskip

\begin{centering}\subsection{Stokes drift}\label{subsec:S4s2}\end{centering}

Concerning the Stokes drift contribution, expansion in powers of a wave-amplitude parameter $\epsilon$ produces the standard approximation \citep{stokes1847theory,phillips1977dynamics,van2018stokes,vanneste2022stokes} to the Stokes velocity
\begin{equation}
\label{eq:EQ23}
\overline{\mathbf{u}}^S=\langle\left(\boldsymbol{\xi}'\boldsymbol{\cdot}\nabla\right)\mathbf{u}'\rangle,
\end{equation}
\noindent with $\mathbf{u}'$ the linear first-order velocity of the wave and $\boldsymbol{\xi}'$ the associated displacement, defined by $\partial_t \boldsymbol{\xi}'=\mathbf{u}'$ and $\langle\boldsymbol{\xi}'\rangle=\langle\mathbf{u}'\rangle=0$.\\
\indent Using formula~\eqref{eq:EQ23}, it can be shown that in the inviscid framework, only the azimuthal component of the drift is nonzero, hence indicating that the poloidal recirculations revealed by experiments can only be due to the steady Eulerian streaming flow or viscous corrections of the Stokes drift \citep{bouvard2017mean}. One of the key points of the present viscous weakly nonlinear analysis is precisely the quantification of these viscous corrections to the Stokes drift and their role in the prediction of the total poloidal Lagrangian mean flow.

\bigskip

\begin{centering}\subsection{Comparison with Bouvard \textit{et al.} (2017) in terms of total Lagrangian mean flow}\label{subsec:S4s3}\end{centering}

\begin{figure}
\centering
\includegraphics[width=1\textwidth]{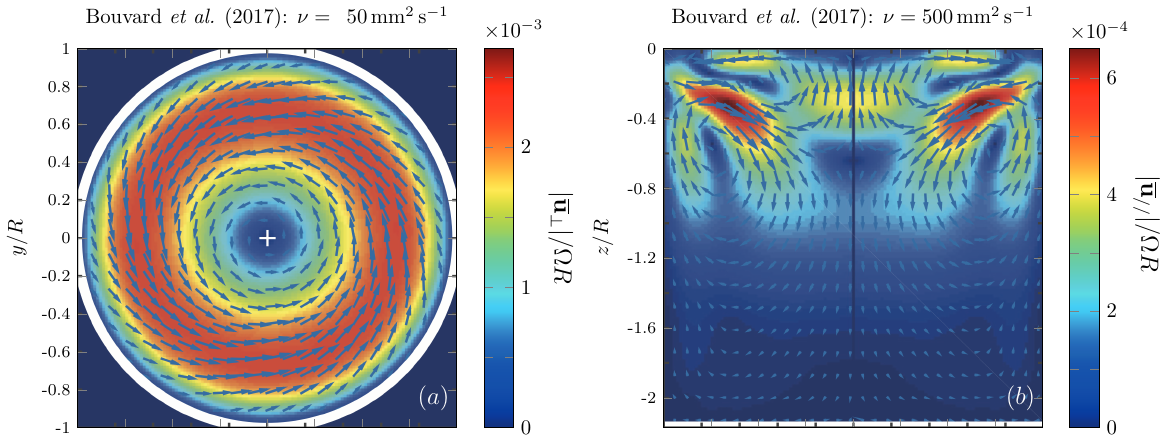}
\includegraphics[width=1\textwidth]{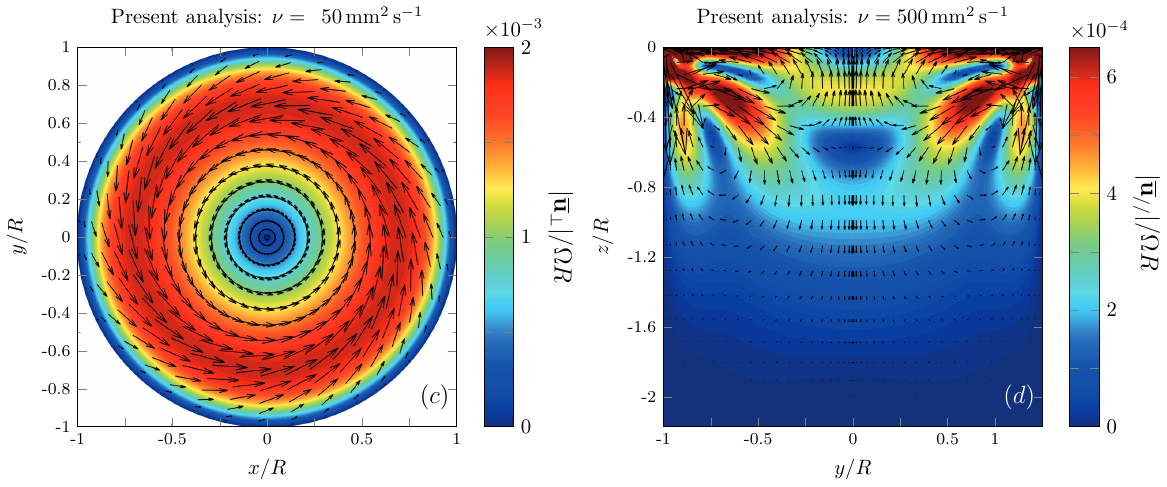}
\caption{(a)-(b) Lagrangian mean flow measured by B17 using stroboscopic PIV. Two different measurements, at the phase $\pi/2$ of the container trajectory, are shown, i.e. (a) in the horizontal plane at a coordinate $z=z_0=-0.23$ below the free surface, and (b) in the vertical plane. These measurements correspond to a driving frequency $\Omega/\omega_1=0.67$, a forcing amplitude $\epsilon=0.057$ and a fluid viscosity $\nu=500\,\text{mm$^2$\,s$^{-1}$}$. Panels (c) and (d) display the same fields shown in (a)-(b), but numerically predicted according to the present viscous weakly nonlinear analysis.}
\label{fig:F4} 
\end{figure}

In comparing our predictions with experiments, let us start from the total Lagrangian mean flow $\overline{\mathbf{u}}$ visualized, by analogy with B17, in the horizontal (toroidal flow) and vertical (poloidal flow) planes. Panels (a) and (b) of figure~\ref{fig:F4} show the reconstructed and symmetrized mean flow fields from B17 for the more viscous fluid with $\nu=500\,\text{mm$^2$\,s$^{-1}$}$.\\
\indent The resemblance between the experimental and the numerical fields, shown in panels (c) and (d), is strong. \textcolor{black}{Note that those panels represent vertical cuts of axisymmetric structures}. As in experiments, the computed toroidal flow appears co-directed with the swirling sloshing wave in each inner point beneath the free surface at a coordinate $z=z_0=-0.23$, tends to zero when approaching the sidewall and reaches a maximum value at approximately $r=0.7$. Analogously, the computed poloidal flow displays two \textcolor{black}{vortex rings}, an upper one with ascending flow along the axis and a lower one with descending flow, separated by a stagnation point located at $z=-0.65$ in experiments and at $z=-0.55$ in the present analysis. The main qualitative difference between the two fields is the strong radial surface flow pointing towards the wall, whose intensity is weaker in the experiments. The intense oblique jets sprouting from the contact line are also well captured. Reproducing the latter aspect is only possible by proper modelling of the sidewall boundary layer and the corner region near the contact line, where the wave amplitude is at its maximum. Appendix~\ref{sec:A1} discusses indeed how variations in the parameters of the slip-length model~\ref{eq:EQ8} lead to significant modifications of the poloidal flow.

\bigskip

\begin{centering}\subsection{Characterization of the individual contributions $\overline{\mathbf{u}}^E$ and $\overline{\mathbf{u}}^S$}\label{subsec:S4s4}\end{centering}

\begin{figure}
\centering
\includegraphics[width=1\textwidth]{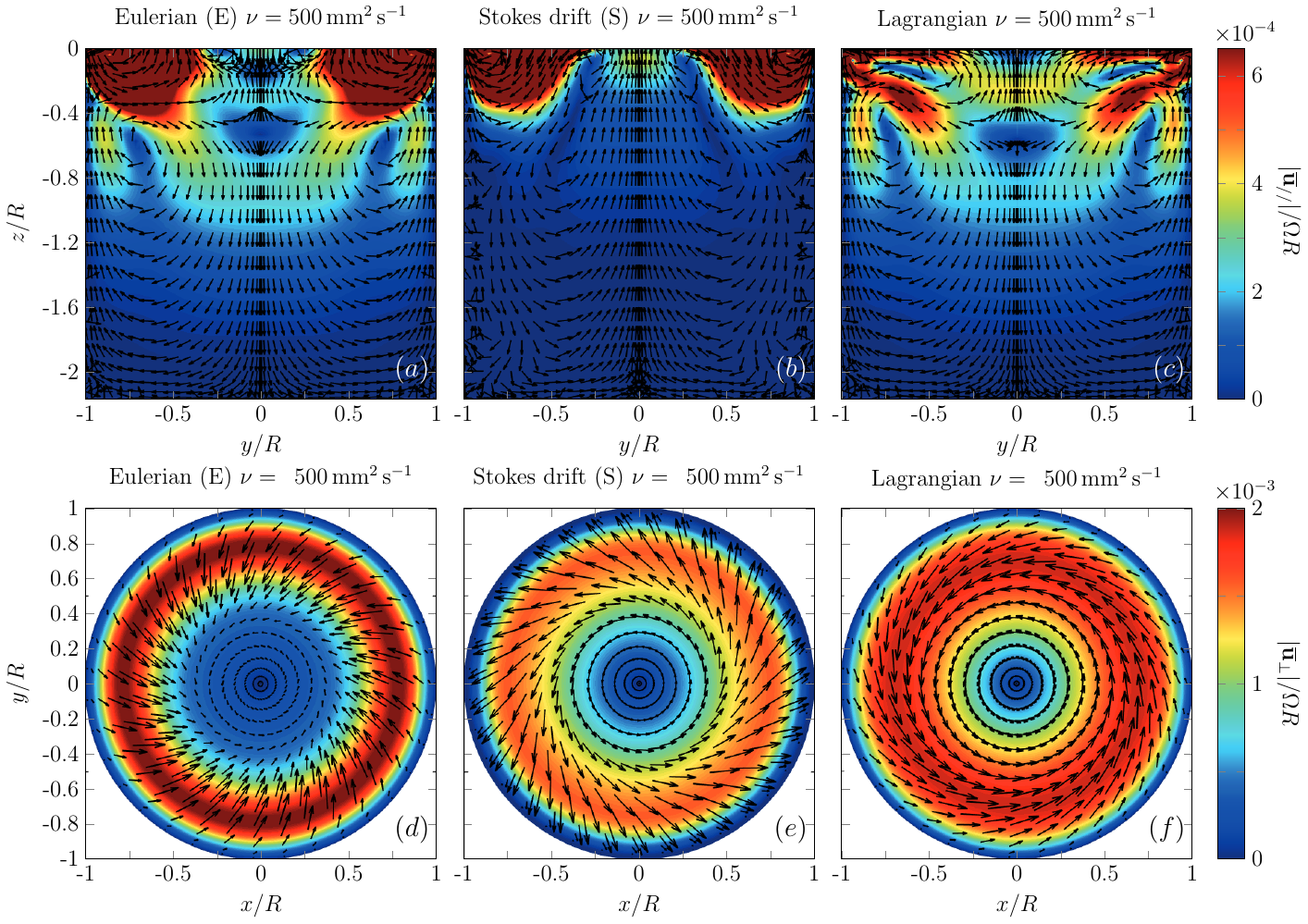}
\caption{(a)-(c) Decomposition of (c) the poloidal Lagrangian mean flow reported in figure~\ref{fig:F4}(d) into (a) its Eulerian component and (b) Stokes drift contribution. (d)-(f) Decomposition of (f) the toroidal Lagrangian mean flow shown in figure~\ref{fig:F4}(c) into (d) its Eulerian component and associated (e) Stokes drift contribution. Note that in (a)-(b) and (d)-(e), the colour bars are saturated to the same maximum value of (c) and (d).}
\label{fig:F5} 
\end{figure}

In contradistinction with the experimental stroboscopic PIV employed by B17, the \textcolor{black}{numerically-based weakly nonlinear approach} here adopted allows us to compute independently and from first principles the individual Eulerian and Stokes drift contributions to the total mean flow in the weakly nonlinear regime. Fields reported in figure~\ref{fig:F4}(c) and (d) are therefore decompose into the $\overline{\mathbf{u}}^E$ and $\overline{\mathbf{u}}^S$ components, which are shown in figure~\ref{fig:F5}.\\
\indent \textcolor{black}{Panel (a) of figure~\ref{fig:F5}} illustrates the poloidal Eulerian flow in the vertical plane. The two jets emanating from the contact line as well as the stagnation point at $z=-0.55$ are intrinsic features of the Eulerian streaming flow. This flow ingredient also shows another stagnation point along the axis at $z=-0.15$, but this feature disappears in the overall flow $\overline{\mathbf{u}}$ (see panel (c)) because of the upward vertical velocity induced by the viscous correction to the Stokes drift in the vertical plane (see panel (b)). Interestingly, whereas the vortex near the contact line is clockwise in the streaming component, it appears anti-clockwise in the Stokes drift component. None of the two fields, taken individually, matches the experimental one, but only their superposition does (panel (c)). This is strong evidence that, at least for highly viscous fluids, radial and axial corrections to the Stokes drift must be retained, as they may be locally or globally of the same order of magnitude as their Eulerian analogous.
\begin{figure}
\centering
\includegraphics[width=1\textwidth]{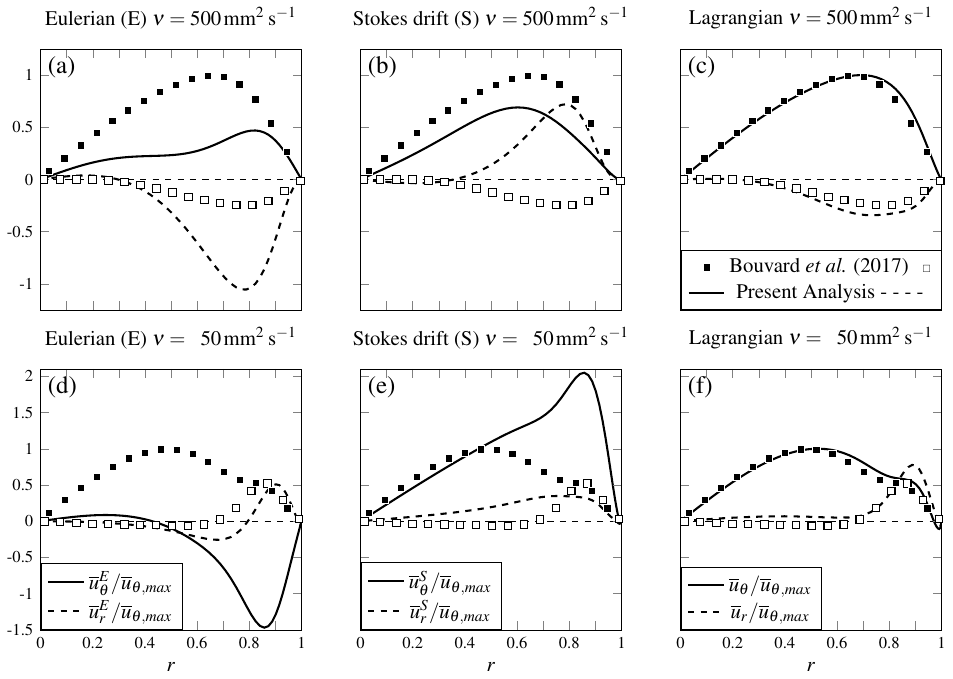}
\caption{Radial and azimuthal velocity profiles of the various mean flow components normalized by the maximum azimuthal Lagrangian velocity (see panel (c) and (f)) and measured at $z=z_0=-0.23$ for $\epsilon=0.057$ and $\Omega/\omega_1=0.67$. (a) Eulerian component, (b) Stokes drift contribution and (c) Lagrangian mean flow computed for $\nu=500\,\text{mm$^2$\,s$^{-1}$}$. (d)-(f) Same but for $\nu=50\,\text{mm$^2$\,s$^{-1}$}$. Black solid and dashed lines correspond, respectively, to the azimuthal and radial velocity profiles numerically computed. Markers correspond instead to the experimental measurements by B17: filled squares for the azimuthal profiles and empty squares for the radial profiles. Note that these measurements give the total Lagrangian mean flow but are here reported in each panel for the sake of comparison with the individual contributions.}
\label{fig:F6} 
\end{figure}\\
\indent Panels (d) and (e) outline the same decomposition but in the horizontal plane for the toroidal flow, \textcolor{black}{showing again the contrast in the Eulerian and Stokes drift components and their simultaneous importance and relevance in their superimposition.} The radial profiles of the radial and azimuthal components normalized by the maximum of the Lagrangian azimuthal component $\overline{u}_{\theta}$ are more clearly displayed in figure~\ref{fig:F6} for $\nu=500\,\text{mm$^2$\,s$^{-1}$}$ (panels (a)-(c)) and $50\,\text{mm$^2$\,s$^{-1}$}$ (panels (d)-(f)). This figure better highlights how only by combining the two contributions $\overline{\mathbf{u}}^E$ and  $\overline{\mathbf{u}}^S$ it is possible to recover the experimental fields. \textcolor{black}{We also observe the contrast between the Eulerian and Stokes drift contributions. While the toroidal Stokes drift at a distance $z=z_0=-0.23$ below the free surface} is co-directed with the container motion, the corresponding Eulerian contribution is counter-directed for the lower viscosity fluid, thus implying return flow, but it is co-directed for the higher viscosity fluids, which is in stark contrast to the traditional picture for weakly viscous fluids \citep{van2018stokes,faltinsen2019inviscid}.
\begin{figure}
\centering
\includegraphics[width=1\textwidth]{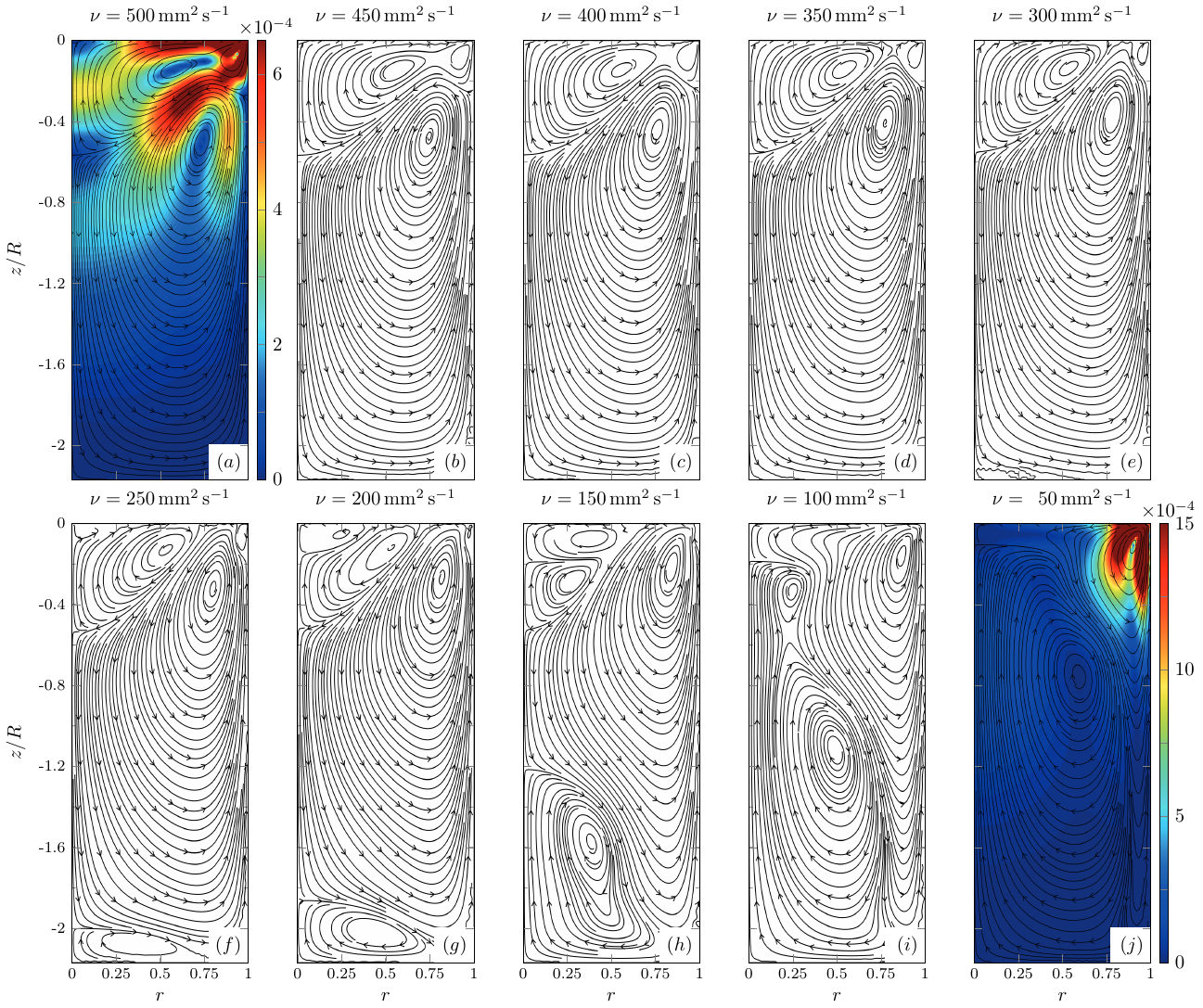}
\caption{Streamlines of the poloidal Lagrangian mean flow shown in the vertical plane for ten different fluid kinematic viscosities ranging from $\nu=500\,\text{mm$^2$\,s$^{-1}$}$ to $\nu=50\,\text{mm$^2$\,s$^{-1}$}$ and computed for $\Omega/\omega_1=0.67$ and $\epsilon=0.057$. This collection of streamlines shows how the poloidal recirculation vortices active in the neighbourhood of the contact line progressively evolve with changes in viscosity. \textcolor{black}{The background colours in panels (a) and (j) show the quantity $|\overline{u}_{//}|/\Omega R$ already displayed in figure~4(d) for $\nu=500\,\text{mm$^2$\,s$^{-1}$}$.}}
\label{fig:F7} 
\end{figure}\\
\indent The radial velocity profiles $\overline{u}_r^E$ and $\overline{u}_r^S$ have opposite signs for $\nu=500\,\text{mm$^2$\,s$^{-1}$}$, but their combination produces an inward radial velocity $\overline{u}_r<0$ that matches the experiments (see panel (c)), with $\overline{u}_r$ nearly zero for $r\rightarrow0$. On the other hand, for $\nu=50\,\text{mm$^2$\,s$^{-1}$}$, the sign of the Eulerian component $\overline{u}_r^E$ is negative for $r<0.8$ and positive for $r>0.8$. The overall profile matches experiments, with $\overline{u}_r$ nearly zero everywhere except for $r\rightarrow 1$. Here, the numerics overestimates the peak of $\overline{u}_r$ at $r=0.95$, but it well captures the local increase of $\overline{u}_{\theta}$ near the wall. Note that the considerations outlined in this section with regards to figure~\ref{fig:F6} only apply to the chosen coordinate $z=z_0=-0.23$, and the relative contribution may differ if selecting other axial coordinates \citep{bouvard2017mean}.

 \bigskip
 
\begin{centering}\subsection{Effect of viscosity on the poloidal recirculation vortex near the contact line}\label{subsec:S4s5}\end{centering}

Taking as reference limits the two fluid viscosity used by B17, we now explore how changes of viscosity $\nu\in\left[50,500\right]\,\text{mm$^2$\,s$^{-1}$}$ modify the poloidal Lagrangian mean flow. To this end, figure~\ref{fig:F7} reproduces the mean flow streamlines computed at different $\nu$ for $\epsilon=0.057$ and $\Omega/\omega_1=0.67$.\\
\indent We have already commented how, at the largest viscosity (panel (a)), there is a first poloidal recirculation vortex rotating clockwise near the contact line, which extends towards the container axis below the free surface. A second vortex, rotating anti-clockwise seats below and invades the whole cell. By decreasing the kinematic viscosity, the upper vortex splits in two (panels (d) and (e)) and a low-intensity vortical structure starts appearing near the bottom wall (panel (f)). By further decreasing $\nu$, the bottom vortex grows in size and migrates upward until it eventually merges with the remaining of the upper clockwise-rotating vortex (panel (i)). The clockwise poloidal recirculation now takes a large portion of the cell, whereas the anti-clockwise vortex is now confined in the neighbourhood of the corner region. In other words, reducing $\nu$ from $500$ to $50\,\text{mm$^2$\,s$^{-1}$}$, produces an inversion in the rotation of the nearest contact line poloidal vortex. Figure~\ref{fig:F7} illustrates how this behaviour, consistent with the experiments described in B17, is progressively revealed by decreasing $\nu$.

\bigskip

\begin{centering}\subsection{Effect of viscosity and driving frequency on the angular velocity at the container axis}\label{subsec:S4s6}\end{centering}

\begin{figure}
\centering
\includegraphics[width=0.8\textwidth]{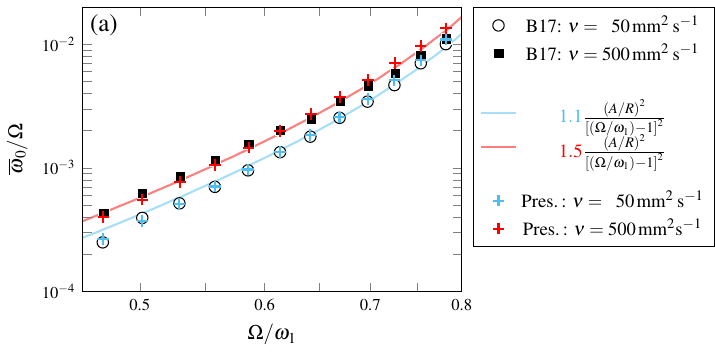}
\includegraphics[width=0.8\textwidth]{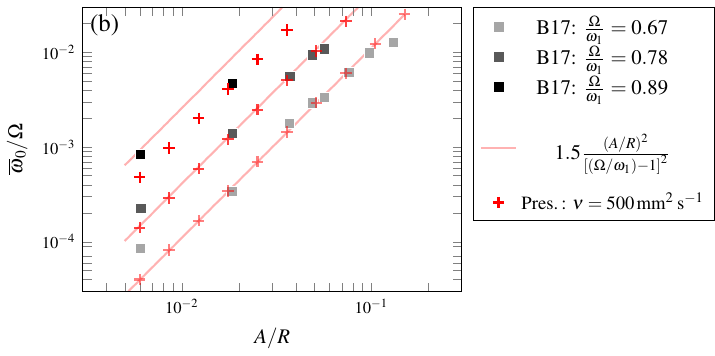}
\caption{(a) Mean angular velocity $\overline{\omega}_0/\Omega=\text{lim}_{r\rightarrow0}\overline{u}_{\theta}\left(r,z=z_0\right)/r\Omega$ at $r=0$, $z=z_0=-0.23$ as a function of the driving frequency $\Omega/\omega_1$ at a fixed forcing amplitude $\epsilon=0.057$, for two fluid viscosities, and (b) as a function of the forcing amplitude $\epsilon=A/R$ at three values of the driving frequency, for $\nu=500\,\text{mm$^2$\,s$^{-1}$}$. In (a), markers correspond to the experiments by B17 and the coloured solid lines show their weakly nonlinear scaling law with a prefactor computed by fitting the measurements. The coloured crosses are instead associated with the present numerical predictions.}
\label{fig:F8} 
\end{figure}

An interesting local measure of the mean flow amplitude as a function of the forcing frequency and amplitude can be extracted by looking at its dominant azimuthal contribution near the centre, i.e. at the angular velocity $\overline{\omega}_0=\lim_{r\rightarrow 0}\overline{u}_{\theta}/r$, e.g. at a depth $z=z_0=-0.23$, which is plotted in figure~\ref{fig:F8}(a) as a function of $\Omega/\omega_1$ for $\epsilon=0.057$ and in figure~\ref{fig:F8}(a) as a function of $\epsilon$ for fixed ratios $\Omega/\omega_1$. In both panels, black markers correspond to the experiments of B17, while the coloured solid lines are given by the asymptotic scaling
\begin{equation}
\label{eq:EQ24}
\frac{\overline{u}}{\Omega R}\sim\frac{\epsilon^2}{\left[\left(\omega_1/\Omega\right)^2-1\right]^2}
\end{equation}
\noindent with a prefactor $K$ fitted from experiments. Note that the scaling above does not depend on the fluid viscosity and is consistent with the calculation by Hutton (1963) \cite{hutton1964fluid}, who only considered the Stokes drift associated with the potential flow solution. However, the Eulerian and Stokes drift contributions scale in the same way and this makes it difficult to discern the two and anticipate which one is more important in determining the total Lagrangian mean flow. While the experiments of Hutton (1963) \cite{hutton1964fluid} were conducted with a very low viscosity fluid, i.e. tap water, for which assuming that the Stokes drift well approximates the solid body rotation near the centre is a reasonable assumption, the experiments by B17 offer a different perspective. We have indeed shown in figure~\ref{fig:F6} that both contributions are important when dealing with moderately or highly viscous fluids. Furthermore, the effect of viscosity appears in the prefactor $K$ required to fit the experimental data in figure~\ref{fig:F8}(a), i.e. $K\simeq1.1\pm0.1$ for the least viscous fluid and $K\simeq1.5\pm0.1$ for the more viscous one.\\
\indent Through the present viscous analysis we can predict the angular velocity in the two cases and we can quantify the effect of viscosity. The coloured crosses reported in panel (a) correspond to the present weakly nonlinear prediction, which gives a good agreement with the experiments and with fitting law by B17, without the need for any fitting parameter.\\
\indent Panel (b) of figure~\ref{fig:F8} shows that our numerical predictions match the scaling law and the experiments at lower driving frequency, but the agreement, as expected, deteriorates approaching the resonant frequency.

\bigskip

\begin{centering}\subsection{Direct comparison with predictions by Faltinsen \& Timokha (2019)}\label{subsec:S4s7}\end{centering}

\begin{figure}
\centering
\includegraphics[width=1\textwidth]{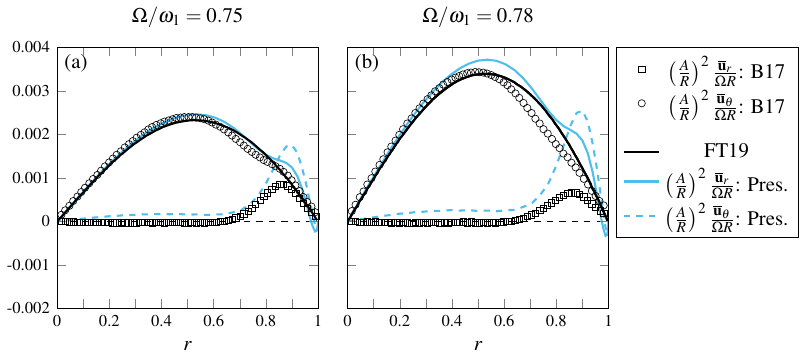}
\caption{The experimentally measured~\onlinecite{bouvard2017mean} (B17) and theoretical (black solid line) \citep{faltinsen2019inviscid} (FT19) liquid mass-transport velocity in the azimuthal (solid circles and solid black line, respectively) and radial (empty circles) directions at $z=z_0=-0.23$, $\Omega/\omega_1=0.75$ (a) and 0.78 (b), and $\epsilon=0.057$. The solid and dashed light blue lines correspond to the azimuthal and radial profiles, respectively, as predicted by the present analysis.}
\label{fig:F9} 
\end{figure}

In this final section, we compare our predictions with the theoretical analysis of Faltinsen \& Timokha (2019) (FT19) \cite{faltinsen2019inviscid} and the experimental profiles of B17.\\
\indent The theory of FT19 predicts the overall azimuthal mass transport as the sum of an azimuthal non-Eulerian mean, i.e. Stokes drift type, and a vortical Eulerian-mean flow, governed by the inviscid Craik-Leibovich equation \citep{craik1976rational}, whose unique non-trivial solution is selected by inputting the information that the mass-transport velocity should tend to zero when approaching the lateral wall. In contradistinction with the present analysis, the model of FT19 is built within the inviscid framework and therefore it cannot predict the axial and radial mean flow velocities, i.e. the poloidal flow. However, while our study assumes a straightforward asymptotic expansion that is not suitable to describe the resonant regime, the study of FT19 is based on the Narimanov-Moiseev multimodal sloshing theory \citep{narimanov1957movement,moiseev1958theory,dodge1965liquid,faltinsen1974nonlinear,narimanov1977,lukovsky1990}. The theory is based on single-mode linear sloshing solution for computing the amplitude parameter and uses a rigorous multiple timescales framework for describing the nonlinear wave dynamics near the primary harmonic resonance \citep{faltinsen2005resonant,faltinsen2016resonant,raynovskyy2020sloshing}. A more accurate quantification of the wave amplitude when approaching the resonant condition $\Omega/\omega_1\rightarrow1$ is important in predicting the intensity of the mean flow. We believe that this is the reason why the azimuthal mass-transport velocity given by our model is in good agreement with FT19 and B17 at $\Omega/\omega_1=0.75$, while this agreement becomes worse at $\Omega/\omega_1=0.78$.\\
\indent Consistently with experiments, the estimated radial velocity shows a peak around $r=0.9$, but its value overestimates the measurements, especially at $\Omega/\omega_1=0.78$, suggesting that nonlinearities in the wave are no longer negligible.

\bigskip

\begin{centering}\section{Conclusions}\label{sec:S5}\end{centering}

In this paper, we built a numerically-based viscous weakly nonlinear analysis capable of reproducing the wave flow and the associated Lagrangian mean flow experimentally observed by Bouvard \textit{et al.} (2017) (B17) \cite{bouvard2017mean} for an orbitally shaken cylinder in nonresonant conditions, i.e. $\Omega/\omega_1<0.8$, and filled with moderately or highly viscous fluids, \textcolor{black}{i.e. $Re=\Omega R^2/\nu\lesssim 1000$}. These predictions include (i) the spatial structure of the first-order wave field and its phase delay with respect to the external driving, which increases with viscosity; (ii) the toroidal (azimuthal) mean flow in the horizontal plane; (iii) the poloidal recirculation vortices mostly active near the contact line and their behaviours as control parameters, such as the fluid viscosity, are varied.\\
\indent The last aspect represents one of the main contributions of this work. Indeed, the description of the poloidal flow, even in the weak streaming regime, represents a challenging task as it requires the modelisation of the viscous sloshing modes and, specifically, the description of the flow dynamics near the contact line, which is essential in determining the spatial structure of the mean recirculation vortices. This was done here numerically by solving the full viscous hydrodynamics problem supplemented with a depth-dependent slip-length model \citep{bongarzone2022numerical}, which was shown to provide a good description of the boundary layer flow near the lateral wall and the resulting second-order weakly nonlinear momentum transfer from the boundary layer region to the fluid bulk, from which the Eulerian mean flow originates.\\
\indent The analysis allowed us to characterize the two contributions to the total Lagrangian mean flow, i.e. the Eulerian mean flow and the Stokes drift, which were not individually accessible in the experiments of B17. The toroidal (azimuthal) mean flow acting in the horizontal plane has been interpreted so far as mainly produced by the Stokes drift, which is traditionally seen as a purely kinematic contribution, with the Eulerian streaming flow only responsible for the poloidal flow in the vertical plane. We showed here that this interpretation fails for moderately or highly viscous sloshing dynamics, for which viscous corrections to the Stokes drift have the same order of magnitude as the Eulerian streaming flow and therefore contribute to defining the structure of the poloidal Lagrangian mean flow. The same holds for the toroidal (azimuthal) component of the Eulerian streaming flow, which may be of the same order of magnitude as the toroidal (azimuthal) Stokes drift. Whereas the toroidal (azimuthal) Stokes drift at a given distance below the free surface is always co-directed with the container motion, we showed that the corresponding Eulerian contribution is largely counter-directed to the container motion for lower fluid viscosities, thus implying return flow \citep{van2018stokes}, but can be co-directed for highly viscous fluids. We also elucidated the effect of viscosity variations in the reorganization of the poloidal vortices near the contact line, whose predicted structure matched fairly well with the experimental one.\\
\indent The numerical analysis developed in this work offers a powerful and efficient tool allowing mean flow predictions without the need of time-stepping computationally expensive three-dimensional direct numerical simulations, as the steady state linear wave and second-order mean flow are here computed at once. The analysis is still limited to the weak streaming limit, as it is based on a straightforward asymptotic expansion that cannot account for nonlinearities in the wave and streaming flow. This precludes one from exploring close-to-resonance conditions or considering low-viscosity fluids for which strong streaming is expected to occur. However, the present results constitute the starting point in the development of \textcolor{black}{rigorous multiple timescale analyses}, through which a system of amplitude equations, coupling the wave dynamics to the mean flow evolution, could be derived in the same spirit of Ref.~\onlinecite{higuera2002coupled}, hence enabling us to revisit prior inviscid analyses \citep{bongarzone2022amplitude,marcotte2023super,marcotte2023swirling} based on the more realistic viscous modes constructed numerically.\\
\indent Lastly, it is known that a free-end edge condition at the contact line is experimentally extremely difficult to reproduce. The flexibility of the numerical tools here proposed would strongly ease accounting for additional effects, such as surface contamination \citep{craik1982drift,martin2006effect,higuera2014nonlinear,perinet2017streaming}, contact angle variation \citep{bongarzone2023revised} or contact line pinning \citep{bongarzone2021relaxation,bongarzone2022sub,bongarzone2024stick}, thus paving a way for a more comprehensive investigation of \textcolor{black}{the influence of these sidewall non-idealities on the resulting steady streaming flow.}








\appendix

\bigskip

\begin{centering}\section{Sensitivity analysis to the value of the slip region penetration depth}\label{sec:A1}\end{centering}

\begin{figure}
\centering
\includegraphics[width=0.875\textwidth]{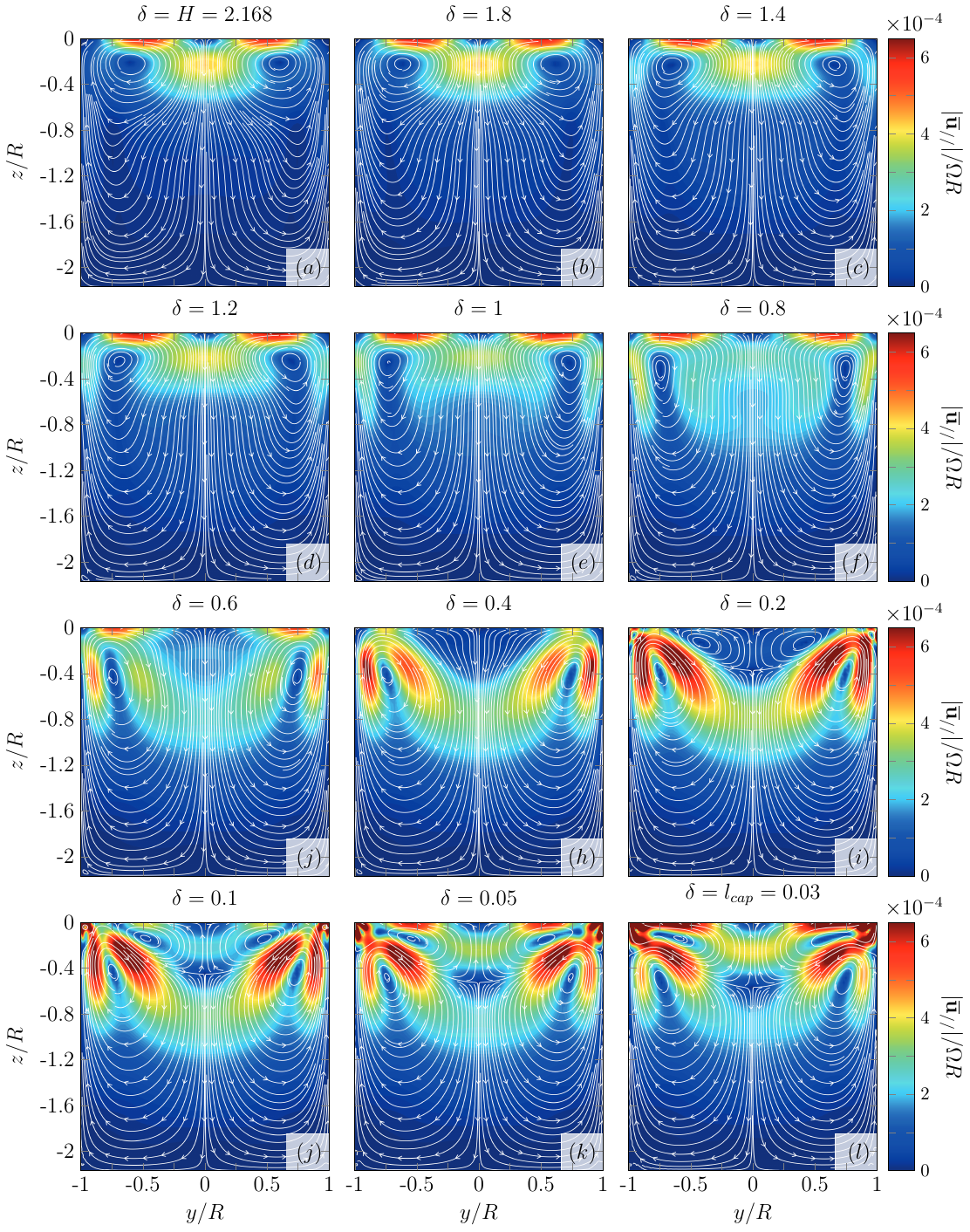}
\caption{Sensitivity analysis of the poloidal Lagrangian mean flow in the vertical plane to variation of the slip region penetration depth $\delta$ adopted in the macroscopic depth-dependent slip-length model~\eqref{eq:EQ8}. Starting from the container depth, $\delta=H=2.168$, the extension of the penetration length is progressively reduced up to the size of the capillary length, $l_cap=0.0305$. The computation is performed at $\Omega/\omega_1=0.67$, $\epsilon=0.057$ and for $\nu=500\,\text{mm$^2$\,s$^{-1}$}$, as in figure~\ref{fig:F4}. The value of $l_{cl}$ and $l_{\delta}$ are set, respectively, to $10^2$ and $10^{-4}$.}
\label{fig:F10} 
\end{figure}

In \S\ref{subsec:S2s3}, we have introduced the depth-dependent slip-length model~\eqref{eq:EQ8}, which has already been employed and validated in \cite{bongarzone2022numerical,bongarzone2024stick} and which shares close similarities with other localized slip conditions commonly employed in the modelization of contact line motions (see for instance \cite{fullana2022simulation}).\\
\indent The model assumes a large value of the slip length $l_s\left(z=0\right)=l_{cl}\sim 10^{2}\div10^{4}$ at the contact line, e.g. $10^3$, and a small value $l_s\left(z=-\delta\right)=l_{\delta}\sim 10^{-4}\div10^{-6}$, e.g. $10^{-5}$ at a distance $\delta$ below the contact line. A large value of $l_{cl}$ is necessary to preserve as much as possible the continuity of vorticity by approaching the contact line from the interface or the lateral wall \citep{miles1990capillary}. Particularly, if the contact line is modelled as a free-end edge condition as in the present study, i.e. $\partial\eta/\partial r=0$, then the value of $l_s$ should tend to infinity. On the other hand, a small value of $l_{\delta}$ is necessary to reproduce the no-slip condition \citep{ting1995boundary} to recover the lateral boundary layer, whose presence is a key feature in the generation of the mean Eulerian streaming flow.\\
\indent If the values of $l_{cl}$ and $l_{\delta}$ are picked within the ranges given above, the numerical solution does not show significant changes. The law adopted here to describe the transition region is exponential as suggested by \citep{ting1995boundary}; different laws, e.g. a linear decay, could affect the final solution, however, we did not investigate this aspect. The fundamental free parameter of the model~\eqref{eq:EQ8} is the slip region penetration depth, $\delta$. We have anticipated in \S\ref{subsec:S2s3} that what mostly matters in choosing the value of this parameter is that $\delta$ should be kept small compared to all the other scales at hand in the problem, i.e. $H$, $R$, capillary length $1/\sqrt{Bo}$, Stokes boundary layer thickness $\sqrt{2/Re}$ or, at most, comparable to the smallest one.\\
\indent In this Appendix, we provide a systematic sensitivity analysis of the poloidal Lagrangian mean flow to significant variations of the parameter $\delta$. Starting from the container depth, $\delta=H=2.168$, the extension of the penetration length is progressively reduced up to the size of the capillary length, $l_{cap}=0.0305$.\\
In the first five panels (a)-(e) of figure~\ref{fig:F10}, the value of $\delta$ is large ($\ge1$) and the sidewall behaves largely as a slip wall. The no-slip condition is exactly enforced at the bottom, but the fluid layer is too deep for the bottom boundary layer to contribute to the mean flow generation, i.e. the wave motion decays exponentially before reaching $z=-H$. In the scenario, the only boundary layer at play is the one below the free surface, where most of the streaming flow generation occurs.\\
\indent For $1<\delta<0.5$ (see panels (f)-(j)), the lateral wall starts to behave as a no-slip wall below the interface and, as a consequence, we observe an intensification of the near-wall mean flow. Only starting from $\delta<0.5$ (panel (h)) we commence retrieving the \textcolor{black}{oblique toroidal jets} sprouting from the contact line region, which becomes progressively more intense by further decreasing $\delta$ to values $<0.2$ (panel (i)-(j)). We note that, at least in this particular case, it is only when approaching the value of the capillary length, i.e. $\delta\approx l_{cap}$ (see panel (k)-(l)) that the structure of the poloidal Lagrangian mean flow is both qualitatively and quantitatively similar to that reported experimentally by \cite{bouvard2017mean}.

\bigskip

\begin{centering}\section{Comparison with other prior works}\label{sec:A2}\end{centering}

\begin{figure}
\centering
\includegraphics[width=0.95\textwidth]{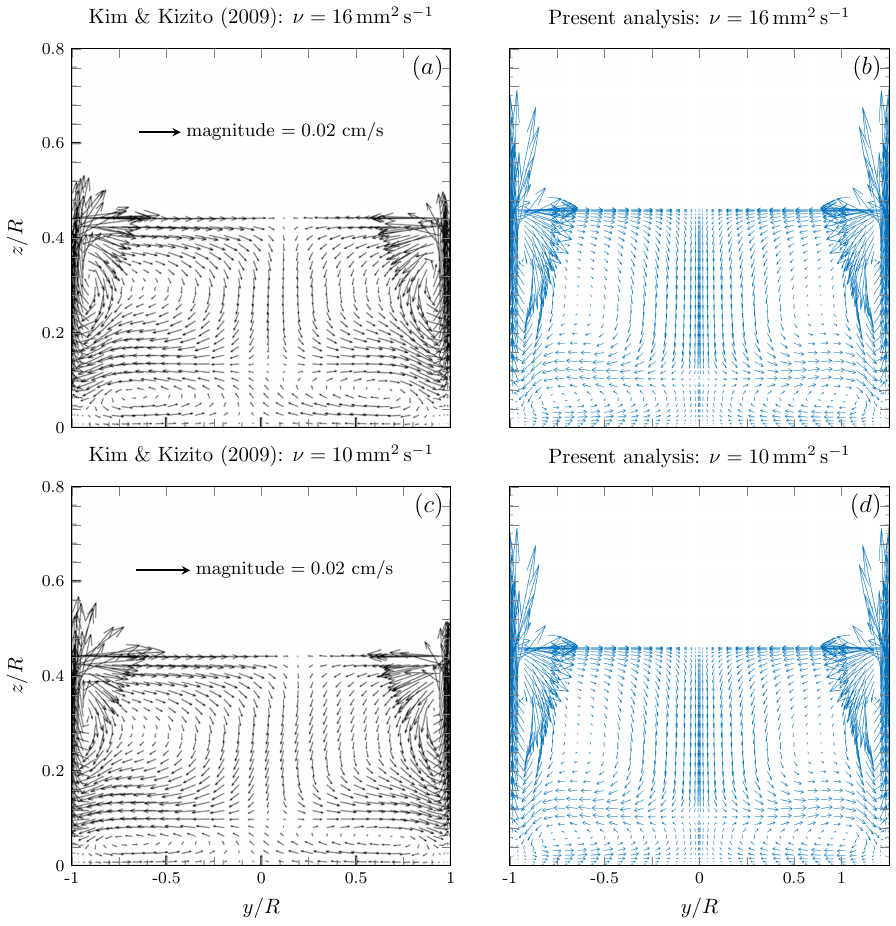}
\caption{(a) Poloidal Lagrangian mean flow, displayed in the vertical plane numerically, computed by \cite{kim2009stirring} for a container of radius $R=43.5\,\text{mm}$ filled to a depth $h=20\,\text{mm}$ with a mixture of water and ethylene glycol liquid of kinematic viscosity $\nu=16\,\text{mm$^2$\,s$^{-1}$}$. The forcing amplitude is also $A=20\,\text{mm}$, corresponding to a large value of $\epsilon$, i.e. $\epsilon=0.46$. The period of the forcing is $1\,\text{s}$, corresponding to $\Omega/\omega_1=0.37$. The scale of the arrows is indicated in the plot. (b) Same as in (a) but according to our \textcolor{black}{weakly nonlinear analysis}. We note that the predicted maximal velocity magnitude is around $0.2\,\text{cm/s}$, which is around one order of magnitude larger than that of \cite{kim2009stirring}. (c)-(d) Same as in (a)-(b) but for $\nu=10\,\text{mm$^2$\,s$^{-1}$}$.}
\label{fig:F11} 
\end{figure}

In the present manuscript, we have used the results by \cite{bouvard2017mean} (B17) as the main reference for comparison with experiments. Except for this seminal work, the sloshing literature lacks comprehensive investigations of the mean flow associated with the orbital motion. In addition, as mentioned by B17, the strong dependence of the poloidal flow with the governing parameters makes difficult direct comparisons with prior works \citep{weheliye2013fluid}.\\
\indent In this appendix, we attempt to apply our viscous weakly nonlinear analysis to the cases considered by \cite{kim2009stirring}. Employing the so-called \textit{Marker-and-Cell} method, they numerically resolved the three-dimensional sloshing problem and computed the wave dynamics as well as the associated mean flow. Their numerical results, which were shown qualitatively similar to those extracted through flow visualisation by homemade experiments, will be used in the following for comparison with our numerical analysis.\\
\indent Panels (a) and (c) of figure~\ref{fig:F11} show the poloidal Lagrangian mean flow calculated by \cite{kim2009stirring} for a container of radius $R=43.5\,\text{mm}$, filled to a depth $h=20\,\text{mm}$ with a mixture of water and ethylene glycol liquid of kinematic viscosity $\nu=16\,\text{mm$^2$\,s$^{-1}$}$, and orbitally forced with a driving period of $1\,\text{s}$ and a forcing amplitude $A=20\,\text{mm}$, which correspond, respectively, to $\Omega/\omega_1=0.37$ and $\epsilon=A/R=0.46$. Specifically, panel (a) is for a fluid with kinematic viscosity $\nu=16\,\text{mm$^2$\,s$^{-1}$}$ and panel (b) for $\nu=10\,\text{mm$^2$\,s$^{-1}$}$. Panels (b) and (d) show the corresponding fields predicted according to our analysis. Qualitatively speaking the agreement is fairly good. Both numerical solutions display the same recirculation regions, i.e. a large anti-clockwise vortex that invades the whole cell and a smaller clockwise vortex flattened to the bottom wall.\\
\indent Nevertheless, from the quantitative point of view the two numerical solutions differ by approximately one order of magnitude, i.e. the magnitude of the arrows reported in panels (b) and (d) is around ten times larger than that of panels (a) and (c). Several explanations are possible: (i) as \cite{kim2009stirring} did not report the fluid properties of the mixture employed in the analysis (except the kinematic viscosity), we have here used the values of water, i.e. $\rho=1000\,\text{kg\,m$^{-3}$}$ and $\gamma=72.5\times10^{-3}\,\text{N\,m$^{-1}$}$; (ii) although the system is driven at a frequency far from resonance at $\Omega/\omega_1=0.37$, the amplitude of the forcing is notably large, i.e. $\epsilon=0.46$, and could therefore violate the validity of the asymptotic analysis; (iii) the fluid viscosity is here five times smaller than the lowest one used in B17, who have cautiously pointed out that the weak streaming assumption is not fully satisfied at $\nu=50\,\text{mm$^{2}$\,s$^{-1}$}$. In this scenario, nonlinearities in the streaming flow, overlooked by the present weakly nonlinear analysis, start perhaps to become important and saturate the mean flow at a lower amplitude, whose value can be instead captured by fully nonlinear direct numerical simulations.

\bibliographystyle{unsrt}
\bibliography{Bibliography}

\begin{thebibliography}{10}

\bibitem{bouvard2017mean}
J.~Bouvard, W.~Herreman, and F.~Moisy.
\newblock Mean mass transport in an orbitally shaken cylindrical container.
\newblock {\em Phys. Rev. Fluids}, 2(8):084801, 2017.

\bibitem{mcdaniel1969effect}
L.~E. McDaniel and E.~G. Bailey.
\newblock Effect of shaking speed and type of closure on shake flask cultures.
\newblock {\em Appl. Microbiol.}, 17(2):286--290, 1969.

\bibitem{wurm2004production}
F.~M. Wurm.
\newblock Production of recombinant protein therapeutics in cultivated
  mammalian cells.
\newblock {\em Nat. Biotechnol.}, 22(11):1393--1398, 2004.

\bibitem{liu2001development}
C.-M. Liu and L.-N. Hong.
\newblock Development of a shaking bioreactor system for animal cell cultures.
\newblock {\em Biochem. Eng. J.}, 7(2):121--125, 2001.

\bibitem{de2004tubespin}
M.~J.~De Jesus, P.~Girard, M.~Bourgeois, G.~Baumgartner, B.~Jacko, H.~Amstutz,
  and F.~M. Wurm.
\newblock Tubespin satellites: a fast track approach for process development
  with animal cells using shaking technology.
\newblock {\em Biochem. Eng. J.}, 17(3):217--223, 2004.

\bibitem{muller2007scalable}
N.~Muller, M.~Derouazi, F.~Van Tilborgh, S.~Wulhfard, D.~L. Hacker, M.~Jordan,
  and F.~M. Wurm.
\newblock Scalable transient gene expression in chinese hamster ovary cells in
  instrumented and non-instrumented cultivation systems.
\newblock {\em Biotechnol. Lett.}, 29(5):703--711, 2007.

\bibitem{handa1989effect}
A.~Handa-Corrigan, A.~N. Emery, and R.~E. Spier.
\newblock Effect of gas—liquid interfaces on the growth of suspended
  mammalian cells: mechanisms of cell damage by bubbles.
\newblock {\em Enzyme Microb. Technol.}, 11(4):230--235, 1989.

\bibitem{kretzmer1991response}
G.~Kretzmer and K.~Sch{\"u}gerl.
\newblock Response of mammalian cells to shear stress.
\newblock {\em Appl. Microbiol. Biotechnol.}, 34(5):613--616, 1991.

\bibitem{papoutsakis1991fluid}
E.~T. Papoutsakis.
\newblock Fluid-mechanical damage of animal cells in bioreactors.
\newblock {\em Trends Biotechnol.}, 9(1):427--437, 1991.

\bibitem{kim2009stirring}
H.~M. Kim and J.~P. Kizito.
\newblock Stirring free surface flows due to horizontal circulatory oscillation
  of a partially filled container.
\newblock {\em Chem. Eng. Commun.}, 196(11):1300--1321, 2009.

\bibitem{buchs2000power1}
J.~B{\"u}chs, U.~Maier, C.~Milbradt, and B.~Zoels.
\newblock Power consumption in shaking flasks on rotary shaking machines: I.
  power consumption measurement in unbaffled flasks at low liquid viscosity.
\newblock {\em Biotechnol. Bioeng.}, 68(6):589--593, 2000.

\bibitem{buchs2000power2}
J.~B{\"u}chs, U.~Maier, C.~Milbradt, and B.~Zoels.
\newblock Power consumption in shaking flasks on rotary shaking machines: Ii.
  nondimensional description of specific power consumption and flow regimes in
  unbaffled flasks at elevated liquid viscosity.
\newblock {\em Biotechnol. Bioeng.}, 68(6):594--601, 2000.

\bibitem{buchs2001introduction}
J.~B{\"u}chs.
\newblock Introduction to advantages and problems of shaken cultures.
\newblock {\em Biochem. Eng. J.}, 7(2):91--98, 2001.

\bibitem{maier2004advances}
U.~Maier, M.~Losen, and J.~B{\"u}chs.
\newblock Advances in understanding and modeling the gas--liquid mass transfer
  in shake flasks.
\newblock {\em Biochem. Eng. J.}, 17(3):155--167, 2004.

\bibitem{muller2005orbital}
N.~Muller, P.~Girard, D.~L. Hacker, M.~Jordan, and F.~M. Wurm.
\newblock Orbital shaker technology for the cultivation of mammalian cells in
  suspension.
\newblock {\em Biotechnol. Bioeng.}, 89(4):400--406, 2005.

\bibitem{micheletti2006fluid}
M.~Micheletti, T.~Barrett, S.~D. Doig, F.~Baganz, M.~S. Levy, J.~M. Woodley,
  and G.~J. Lye.
\newblock Fluid mixing in shaken bioreactors: Implications for scale-up
  predictions from microlitre-scale microbial and mammalian cell cultures.
\newblock {\em Chem. Eng. Sci.}, 61(9):2939--2949, 2006.

\bibitem{zhang2009efficient}
X.~Zhang, C.~B{\"u}rki, M.~Stettler, D.~De Sanctis, M.~Perrone, M.~Discacciati,
  N.~Parolini, M.~DeJesus, D.~L. Hacker, A.~Quarteroni, and F.~M. Wurm.
\newblock Efficient oxygen transfer by surface aeration in shaken cylindrical
  containers for mammalian cell cultivation at volumetric scales up to 1000 l.
\newblock {\em Biochem. Eng. J.}, 45(1):41--47, 2009.

\bibitem{tissot2010determination}
S.~Tissot, M.~Farhat, D.~L. Hacker, T.~Anderlei, M.~K{\"u}hner, C.~Comninellis,
  and F.~M. Wurm.
\newblock Determination of a scale-up factor from mixing time studies in
  orbitally shaken bioreactors.
\newblock {\em Biochem. Eng. J.}, 52(2-3):181--186, 2010.

\bibitem{tan2011measurement}
R.-K. Tan, W.~Eberhard, and J.~B{\"u}chs.
\newblock Measurement and characterization of mixing time in shake flasks.
\newblock {\em Chem. Eng. Sci.}, 66(3):440--447, 2011.

\bibitem{tissot2011efficient}
S.~Tissot, A.~Oberbek, M.~Reclari, M.~Dreyer, D.~L. Hacker, L.~Baldi,
  M.~Farhat, and F.~M. Wurm.
\newblock Efficient and reproducible mammalian cell bioprocesses without probes
  and controllers?
\newblock {\em New Biotechnol.}, 28(4):382--390, 2011.

\bibitem{klockner2012advances}
W.~Kl{\"o}ckner and J.~B{\"u}chs.
\newblock Advances in shaking technologies.
\newblock {\em Trends Biotechnol.}, 30(6):307--314, 2012.

\bibitem{abramson1966dynamic}
H.~N. Abramson.
\newblock {\em The dynamic behavior of liquids in moving containers, with
  applications to space vehicle technology}.
\newblock \textit{NASA Tech. Rep.} SP-106. NASA, Washington, 1966.

\bibitem{abramson1966some}
H.~N. Abramson, W.-H. Chu, and D.~D. Kana.
\newblock {\em Some studies of nonlinear lateral sloshing in rigid containers}.
\newblock \textit{NASA Contractor Rep.} NASA CR-375. NASA, 1966.

\bibitem{chu1968subharmonic}
W.-H. Chu.
\newblock Subharmonic oscillations in an arbitrary tank resulting from axial
  excitation.
\newblock {\em Trans. ASME J. Appl. Mech.}, 35:148--154, 1968.

\bibitem{royon2007liquid}
A.~Royon-Lebeaud, E.~J. Hopfinger, and A.~Cartellier.
\newblock Liquid sloshing and wave breaking in circular and square-base
  cylindrical containers.
\newblock {\em J.~Fluid Mech.}, 577:467--494, 2007.

\bibitem{hopfinger2009liquid}
E.~J. Hopfinger and V.~Baumbach.
\newblock Liquid sloshing in cylindrical fuel tanks.
\newblock {\em Prog.~Propul.~Phys.}, 1:279--292, 2009.

\bibitem{reclari2013hydrodynamics}
M.~Reclari.
\newblock Hydrodynamics of orbital shaken bioreactors.
\newblock Technical report, EPFL, 2013.

\bibitem{reclari2014surface}
M.~Reclari, M.~Dreyer, S.~Tissot, D.~Obreschkow, F.~M. Wurm, and M.~Farhat.
\newblock Surface wave dynamics in orbital shaken cylindrical containers.
\newblock {\em Phys. Fluids}, 26(5):052104, 2014.

\bibitem{moisy2018counter}
F.~Moisy, J.~Bouvard, and W.~Herreman.
\newblock Counter-rotation in an orbitally shaken glass of beer.
\newblock {\em EPL}, 122(3):34002, 2018.

\bibitem{faltinsen2016resonant}
O.~M. Faltinsen, I.~A. Lukovsky, and A.~N. Timokha.
\newblock Resonant sloshing in an upright annular tank.
\newblock {\em J.~Fluid Mech.}, 804:608--645, 2016.

\bibitem{timokha2017}
A.~N. Timokha and I.~Raynovskyy.
\newblock The damped sloshing in an upright circular tank due to an orbital
  forcing.
\newblock {\em Dopov.~Nats.~Akad.~Mauk.~Ukr.}, 10:48--53, 2017.

\bibitem{raynovskyy2018steady}
I.~Raynovskyy and A.~N. Timokha.
\newblock Steady-state resonant sloshing in an upright cylindrical container
  performing a circular orbital motion.
\newblock {\em Math. Probl. Eng.}, 2018.

\bibitem{raynovskyy2018damped}
I.~A. Raynovskyy and A.~N. Timokha.
\newblock Damped steady-state resonant sloshing in a circular base container.
\newblock {\em Fluid Dyn. Res.}, 50(4):045502, 2018.

\bibitem{raynovskyy2020sloshing}
I.~A. Raynovskyy and A.~N. Timokha.
\newblock {\em Sloshing in Upright Circular Containers: Theory, Analytical
  Solutions, and Applications}.
\newblock CRC Press, 2020.

\bibitem{horstmann2020linear}
G.~M. Horstmann, W.~Herreman, and T.~Weier.
\newblock Linear damped interfacial wave theory for an orbitally shaken upright
  circular cylinder.
\newblock {\em J.~Fluid Mech.}, 891, 2020.

\bibitem{horstmann2021formation}
G.~M. Horstmann, S.~Anders, D.~H. Kelley, and T.~Weier.
\newblock Formation of spiral waves in cylindrical containers under orbital
  excitation.
\newblock {\em J.~Fluid Mech.}, 925, 2021.

\bibitem{chen2023mechanism}
B.-F. Chen, C.-H. Wu, and O.~M. Faltinsen.
\newblock The mechanism of switching direction of swirling sloshing waves.
\newblock {\em J.~Fluid Mech.}, 954:A2, 2023.

\bibitem{bongarzone2022amplitude}
A.~Bongarzone, M.~Guido, and F.~Gallaire.
\newblock An amplitude equation modelling the double-crest swirling in
  orbital-shaken cylindrical containers.
\newblock {\em J.~Fluid Mech.}, 943:A28, 2022.

\bibitem{marcotte2023super}
A.~Marcotte, F.~Gallaire, and A.~Bongarzone.
\newblock Super-harmonically resonant swirling waves in longitudinally forced
  circular cylinders.
\newblock {\em J. Fluid Mech.}, 966:A41, 2023.

\bibitem{marcotte2023swirling}
A.~Marcotte, F.~Gallaire, and A.~Bongarzone.
\newblock Swirling against the forcing: Evidence of stable counterdirected
  sloshing waves in orbital-shaken reservoirs.
\newblock {\em Phys. Rev. Fluids}, 8:084802, 2023.

\bibitem{Lamb32}
H.~Lamb.
\newblock {\em Hydrodynamics}.
\newblock Cambridge university press, 1993.

\bibitem{faltinsen2005liquid}
O.~M. Faltinsen and A.~N. Timokha.
\newblock {\em Sloshing}.
\newblock Cambridge University Press, 2009.

\bibitem{ibrahim2009liquid}
R.~A. Ibrahim.
\newblock {\em Liquid sloshing dynamics: theory and applications}.
\newblock Cambridge University Press, 2005.

\bibitem{hutton1963inv}
R.~E. Hutton.
\newblock {\em An investigation of nonlinear, nonplanar oscillations of fluid
  in cylindrical container}.
\newblock \textit{NASA Tech. Rep.} NASA; D-1870, 1963.

\bibitem{stokes1847theory}
G.~G. Stokes.
\newblock On the theory of oscillatory waves.
\newblock {\em Trans. Cam. Philos. Soc.}, 8:441--455, 1847.

\bibitem{van2018stokes}
T.~S. Van~Den Bremer and {\O}.~Breivik.
\newblock Stokes drift.
\newblock {\em Philos. Trans. Royal Soc. A PHILOS T R SOC A}, 376(2111), 2018.

\bibitem{vanneste2022stokes}
J.~Vanneste and W.~R. Young.
\newblock Stokes drift and its discontents.
\newblock {\em Phil. Trans. Royal Soc. A}, 380(2225):20210032, 2022.

\bibitem{batchelor1967introduction}
G.~K. Batchelor.
\newblock {\em An introduction to fluid dynamics}.
\newblock Cambridge university press, 1967.

\bibitem{craik1982drift}
A.~D.~D. Craik.
\newblock The drift velocity of water waves.
\newblock {\em J.~Fluid Mech.}, 116:187--205, 1982.

\bibitem{craik1976rational}
A.~D.~D. Craik and S.~Leibovich.
\newblock A rational model for langmuir circulations.
\newblock {\em J.~Fluid Mech.}, 73(3):401--426, 1976.

\bibitem{perinet2017streaming}
N.~P{\'e}rinet, P.~Guti{\'e}rrez, H.~Urra, N.~Mujica, and L.~Gordillo.
\newblock Streaming patterns in faraday waves.
\newblock {\em J. Fluid Mech.}, 819:285--310, 2017.

\bibitem{higuera2005dynamics}
M.~Higuera, E.~Knobloch, and J.~M. Vega.
\newblock Dynamics of nearly inviscid faraday waves in almost circular
  containers.
\newblock {\em Phys. D: Nonlin. Phen.}, 201(1-2):83--120, 2005.

\bibitem{riley2001steady}
N.~Riley.
\newblock Steady streaming.
\newblock {\em Ann. Rev. Fluid Mech.}, 33(1):43--65, 2001.

\bibitem{weheliye2013fluid}
W.~Weheliye, M.~Yianneskis, and A.~Ducci.
\newblock On the fluid dynamics of shaken bioreactors—flow characterization
  and transition.
\newblock {\em AIChe}, 59(1):334--344, 2013.

\bibitem{faltinsen2019inviscid}
Odd~M Faltinsen and Alexander~N Timokha.
\newblock An inviscid analysis of the prandtl azimuthal mass transport during
  swirl-type sloshing.
\newblock {\em J.~Fluid Mech.}, 865:884--903, 2019.

\bibitem{hutton1964fluid}
R.~E. Hutton.
\newblock Fluid-particle motion during rotary sloshing.
\newblock {\em Trans. ASME J. Appl. Mech.}, 31(1):145--153, 1964.

\bibitem{nicolas2003three}
J.~A. Nicol{\'a}s and J.~M. Vega.
\newblock Three-dimensional streaming flows driven by oscillatory boundary
  layers.
\newblock {\em Fluid Dyn. Res.}, 32(4):119--139, 2003.

\bibitem{Huh71}
C.~Huh and L.~E. Scriven.
\newblock Hydrodynamic model of steady movement of a solid/liquid/fluid contact
  line.
\newblock {\em J. Colloid Interface Sci.}, 35(1):85--101, 1971.

\bibitem{Davis1974}
S.~H. Davis.
\newblock On the motion of a fluid-fluid interface along a solid surface.
\newblock {\em J.~Fluid Mech.}, 65(1):71--95, 1974.

\bibitem{navier1823memoire}
C.~L. M.~H. Navier.
\newblock M{\'e}moire sur les lois du mouvement des fluides.
\newblock {\em M{\'e}m. Acad. R. des Sci. Inst. France}, 6(1823):389--440,
  1823.

\bibitem{Lauga2007}
Eric Lauga, Michael Brenner, and Howard Stone.
\newblock {\em Microfluidics: The No-Slip Boundary Condition}, pages
  1219--1240.
\newblock Springer Berlin Heidelberg, Berlin, Heidelberg, 2007.

\bibitem{Viola2018b}
F.~Viola and F.~Gallaire.
\newblock Theoretical framework to analyze the combined effect of surface
  tension and viscosity on the damping rate of sloshing waves.
\newblock {\em Phys. Rev. Fluids}, 3(9):094801, 2018.

\bibitem{miles1990capillary}
J.~W. Miles.
\newblock Capillary-viscous forcing of surface waves.
\newblock {\em J.~Fluid Mech.}, 219:635--646, 1990.

\bibitem{ting1995boundary}
C.~L. Ting and M.~Perlin.
\newblock Boundary conditions in the vicinity of the contact line at a
  vertically oscillating upright plate: an experimental investigation.
\newblock {\em J.~Fluid Mech.}, 295:263--300, 1995.

\bibitem{bongarzone2022numerical}
A.~Bongarzone and F.~Gallaire.
\newblock Numerical estimate of the viscous damping of capillary-gravity waves:
  A macroscopic depth-dependent slip-length model, 2022.

\bibitem{bongarzone2024stick}
A.~Bongarzone and F.~Gallaire.
\newblock Stick-slip-to-stick transition of liquid oscillations in a u-shaped
  tube.
\newblock {\em Phys. Rev. Fluids}, 9(3):034401, 2024.

\bibitem{bongarzone2021relaxation}
A.~Bongarzone, F.~Viola, and F.~Gallaire.
\newblock Relaxation of capillary-gravity waves due to contact line
  nonlinearity: A projection method.
\newblock {\em Chaos}, 31(12):123124, 2021.

\bibitem{bongarzone2022sub}
A.~Bongarzone, F.~Viola, S.~Camarri, and F.~Gallaire.
\newblock Subharmonic parametric instability in nearly brimful circular
  cylinders: a weakly nonlinear analysis.
\newblock {\em J.~Fluid Mech.}, 947:A24, 2022.

\bibitem{Viola2016a}
F.~Viola, C.~Arratia, and F.~Gallaire.
\newblock Mode selection in trailing vortices: harmonic response of the
  non-parallel {B}atchelor vortex.
\newblock {\em J.~Fluid Mech.}, 790:523--552, 2016.

\bibitem{heinrichs2004spectral}
W.~Heinrichs.
\newblock Spectral collocation schemes on the unit disc.
\newblock {\em J.~Comput. Phys.}, 199(1):66--86, 2004.

\bibitem{canuto2007spectral}
C.~Canuto, M.~Y. Hussaini, A.~Quarteroni, and T.~A. Zang.
\newblock {\em Spectral methods: evolution to complex geometries and
  applications to fluid dynamics}.
\newblock Springer Science \& Business Media, 2007.

\bibitem{sommariva2013fast}
A.~Sommariva.
\newblock Fast construction of fej{\'e}r and clenshaw--curtis rules for general
  weight functions.
\newblock {\em Comput. Math. Appl.}, 65(4):682--693, 2013.

\bibitem{Viola2018a}
F.~Viola, P.-T. Brun, and F.~Gallaire.
\newblock Capillary hysteresis in sloshing dynamics: a weakly nonlinear
  analysis.
\newblock {\em J.~Fluid Mech.}, 837:788--818, 2018.

\bibitem{phillips1977dynamics}
O.~M. Phillips.
\newblock {\em The dynamics of the upper ocean}.
\newblock Cambridge University Press, 1977.

\bibitem{narimanov1957movement}
G.~S. Narimanov.
\newblock Movement of a tank partly filled by a fluid: the taking into account
  of non-smallness of amplitude.
\newblock {\em Prikl.~Math.~Mech. (in Russian)}, 21:513--524, 1957.

\bibitem{moiseev1958theory}
N.~N. Moiseev.
\newblock On the theory of nonlinear vibrations of a liquid of finite volume.
\newblock {\em J. Appl. Math. Mech.}, 22(5):860--872, 1958.

\bibitem{dodge1965liquid}
F.~T. Dodge, D.~D. Kana, and H.~N. Abramson.
\newblock Liquid surface oscillations in longitudinally excited rigid
  cylindrical containers.
\newblock {\em AIAA}, 3(4):685--695, 1965.

\bibitem{faltinsen1974nonlinear}
O.~M. Faltinsen.
\newblock A nonlinear theory of sloshing in rectangular tanks.
\newblock {\em J. Sh. Res.}, 18(04):224--241, 1974.

\bibitem{narimanov1977}
G.~S. Narimanov, L.~V. Dokuchaev, and I.~A. Lukovsky.
\newblock Nonlinear dynamics of flying apparatus with liquid. moscow:
  Mashinostroenie.
\newblock {\em (in Russian)}, 1977.

\bibitem{lukovsky1990}
I.~A. Lukovsky.
\newblock Introduction to nonlinear dynamics of a solid body with a cavity
  including a liquid.
\newblock {\em Kiev: Naukova dumka (in Russian)}, 1990.

\bibitem{faltinsen2005resonant}
O.~M. Faltinsen, O.~F. Rognebakke, and A.~N. Timokha.
\newblock Resonant three-dimensional nonlinear sloshing in a square-base basin.
  part 2. effect of higher modes.
\newblock {\em J.~Fluid Mech.}, 523:199--218, 2005.

\bibitem{higuera2002coupled}
M.~Higuera, J.~M. Vega, and E.~Knobloch.
\newblock Coupled amplitude-streaming flow equations for nearly inviscid
  faraday waves in small aspect ratio containers.
\newblock {\em J. Nonlinear Sci.}, 12:505--551, 2002.

\bibitem{martin2006effect}
E.~Mart{\'\i}n and J.~M. Vega.
\newblock The effect of surface contamination on the drift instability of
  standing faraday waves.
\newblock {\em J. Fluid Mech.}, 546:203--225, 2006.

\bibitem{higuera2014nonlinear}
M.~Higuera, J.~Porter, F.~Varas, and J.~M. Vega.
\newblock Nonlinear dynamics of confined liquid systems with interfaces subject
  to forced vibrations.
\newblock {\em Adv. Colloid Interface Sci.}, 206:106--115, 2014.

\bibitem{bongarzone2023revised}
A.~Bongarzone, B.~Jouron, F.~Viola, and F.~Gallaire.
\newblock A revised gap-averaged floquet analysis of faraday waves in hele-shaw
  cells.
\newblock {\em J.~Fluid Mech.}, 977:A45, 2023.

\bibitem{fullana2022simulation}
T.~Fullana.
\newblock {\em Simulation and optimization of complex phenomena in multiphase
  flows}.
\newblock PhD thesis, Sorbonne Universit{\'e}, 2022.

\end{thebibliography}

\end{document}